\documentclass[11pt]{article}
\usepackage[margin=1in]{geometry}
\usepackage[T1]{fontenc}
\usepackage[utf8]{inputenc}
\usepackage{lmodern}
\usepackage{amsmath,amssymb,amsthm,mathtools}
\usepackage{microtype}
\usepackage{xcolor}
\usepackage{tikz}
\usepackage[numbers,sort&compress]{natbib}
\usepackage{authblk}
\usetikzlibrary{arrows.meta}
\usepackage[colorlinks=true,linkcolor=blue!50!black,urlcolor=blue!55!black]{hyperref}
\newtheorem{theorem}{Theorem}
\newtheorem{lemma}{Lemma}
\newtheorem{proposition}{Proposition}
\newtheorem{corollary}{Corollary}
\theoremstyle{definition}
\newtheorem{definition}{Definition}
\DeclareMathOperator{\ran}{ran}
\DeclareMathOperator{\tr}{tr}
\DeclareMathOperator{\arcosh}{arcosh}
\DeclareMathOperator{\arsinh}{arsinh}
\newcommand{\HH}{\mathcal H}
\newcommand{\EE}{\mathcal E}
\newcommand{\TT}{\mathbb T}
\newcommand{\one}{\mathbf 1}
\newcommand{\phys}{\mathrm{phys}}
\newcommand{\refscale}{\mathrm{ref}}

\title{Improved Digitization Bounds for U(1) Lattice Gauge Theory via Semiclassical Analysis}
\author[1,2]{Di Fang}
\author[1,2,3]{Yu Tong}
\affil[1]{Department of Mathematics, Duke University}
\affil[2]{Duke Quantum Center, Duke University}
\affil[3]{Department of Electrical and Computer Engineering, Duke University}
\date{}

\begin{document}
\maketitle

\begin{abstract}
    Simulating lattice gauge theories on a digital quantum computer or using tensor network methods requires a finite-dimensional approximation of the gauge field, a process often called digitization. Controlling the resulting error is essential for assessing the accuracy and resource requirements of these simulations. In this work, we derive a digitization error bound for low-lying eigenstates of the Kogut--Susskind Hamiltonian for pure U(1) gauge theory and translate it into a bound on time-evolution error. Compared with existing bounds, our result achieves exponentially improved dependence on the inverse lattice spacing $1/a$ and evolution time $T$, together with improved dependence on the inverse coupling $1/g$. The key idea is to identify a small parameter in the Kogut--Susskind Hamiltonian and apply tools from semiclassical analysis. Specifically, we adapt Agmon's method to establish exponential decay of the error with the truncation threshold. These results sharpen estimates of the local Hilbert space dimension and qubit resources needed to simulate gauge theories near the continuum limit, at weak coupling, and over long times.
\end{abstract}

\section{Introduction}

Gauge theories describe the fundamental interactions of elementary
particles, and their real-time dynamics are an important target for quantum
simulation~\cite{bauer2023fundamental}. A standard starting point is the
Kogut--Susskind Hamiltonian formulation, which discretizes space while
retaining continuous time~\cite{kogut1975hamiltonian}. Spatial
discretization alone, however, does not produce a finite-dimensional
quantum system needed for a digital quantum computer or even many classical simulation methods such as tensor networks, since each link still carries an infinite-dimensional gauge-field
Hilbert space. 
We refer to the replacement of each gauge link’s infinite-dimensional Hilbert space by a finite-dimensional approximation as \emph{digitization}.
For lattice gauge theories with compact Lie gauge groups, representation theory provides natural finite-dimensional truncations of the link Hilbert spaces.
As an example, for U(1) gauge theory, the electric field on a link has
integer eigenvalues, making it natural to impose an integer cutoff $\Lambda\geq0$, keeping only
the electric eigenstates corresponding to eigenvalues between $-\Lambda$ and $\Lambda$. 
A natural question is then, how do we quantify the error that results from this truncation?

Considerable effort has been devoted to understanding this approximation error, which we refer to as the \emph{digitization error}, and to finding economical field representations. For example, digitization error has been bounded in terms of the energy using Markov's inequality~\cite{jordan2011quantum} and analyzed using sampling theory~\cite{macridin2018digital,klco2019digitization,macridin2022bosonic}. Further analytical and numerical studies of this error include Refs.~\cite{hackett2019digitizing,haase2021resource,davoudi2021search,tong2022provably,hanada2023estimating,calliari2026field,yang2026tightening,ciavarella2026truncation,peng2025quantum,shaw2020quantum}. Methods have also been developed to optimize truncation schemes by choosing alternative bases or suitable discrete subgroups of the gauge group~\cite{alexandru2019gluon,haase2021resource,bauer2023efficient}.

A mathematically rigorous framework to analyze this error was proposed in Ref.~\cite{tong2022provably}, which obtained bounds for a broad class of gauge theories and bosonic systems. 
This framework addresses evolution starting from states with bounded local quantum numbers.
These bounds achieve polylogarithmic dependence on the inverse precision $\epsilon^{-1}$, going beyond the polynomial dependence achieved through energy-based bounds such as the one in Ref.~\cite{jordan2011quantum}. However, in the setting of lattice gauge theories, the bounds in Ref.~\cite{tong2022provably} scale linearly in $T/(ag^2)$, where $T$ is the simulation time, $a$ is the lattice spacing, and $g$ is the coupling that we will introduce later. These bounds therefore become unfavorable for long simulation times, small lattice spacings needed to approach the continuum limit, or weak coupling in the magnetically dominated regime.

We will obtain greatly improved dependence on these parameters for simulation in a low-energy subspace. For concreteness we will focus on 
pure U(1) gauge theory without matter or background charges
on a hypercubic lattice in $d\geq2$ spatial dimensions. Our methods extend beyond this specific setting. Let $a>0$ be the
lattice spacing, $g>0$ the dimensionless coupling, and $\mathsf L$ and
$\mathsf P$ the sets of oriented links and plaquettes. For a link $\ell$,
let $\theta_\ell\in\mathbb R/(2\pi\mathbb Z)$ be its angle and
$E_\ell=-i\partial_{\theta_\ell}$ its electric operator. For a plaquette
$p$, let $\theta_p$ be the oriented sum of its boundary link angles.
The Hamiltonian is
\begin{equation}\label{eq:intro-hamiltonian}
 H_a=\frac1a\left[
       \frac{g^2}{2}\sum_{\ell\in\mathsf L}E_\ell^2
       +\frac1{g^2}\sum_{p\in\mathsf P}(1-\cos\theta_p)\right].
\end{equation}

The key observation in this work is that the small coupling $g$ results in a mathematical structure that is naturally suited for \emph{semiclassical analysis}.
We set $h=g^2$, and write
$\theta=(\theta_\ell)_{\ell\in\mathsf L}$. Let
$\Delta_\theta=\sum_{\ell\in\mathsf L}\partial_{\theta_\ell}^2$ be the
Laplacian in the link angles. We then rescale the Hamiltonian to be
\begin{equation}\label{eq:intro-semiclassical}
 K:=ahH_a=-\frac{h^2}{2}\Delta_\theta+V(\theta),
 \quad V(\theta):=\sum_{p\in\mathsf P}(1-\cos\theta_p).
\end{equation}
Weak coupling $g\to 0$ therefore corresponds to a semiclassical limit $h\to0$.
Near a minimum of the magnetic potential, balancing the kinetic energy
against quadratic confinement suggests an angular width of order
$\sqrt h$ in the confined directions \cite{simon1983semiclassical}. The corresponding Fourier width,
and hence the scale of electric numbers, is $h^{-1/2}=g^{-1}$. This
motivates a localization estimate on the scale $g^{-1}$. We will use tools from semiclassical analysis to control the weight beyond this Fourier width.

We assume that the initial state is contained in an energy window above the
interacting vacuum. More precisely, we let $E_0(a)$ be the ground energy of $H_a$ in the
gauge-invariant Hilbert space, let $\EE\geq0$ be a finite excitation
window, and assume the initial state is a superposition of physical (i.e., obeying Gauss's law) eigenstates
with energies in $[E_0(a),E_0(a)+\EE]$. Therefore, the initial state assumption in this work is different from that in Ref.~\cite{tong2022provably}.

Let $P_\Lambda$ be the orthogonal projection onto states whose electric
numbers all lie in $[-\Lambda,\Lambda]$, and define
\[
H_{a,\Lambda}
=\left.P_\Lambda H_aP_\Lambda\right|_{\ran P_\Lambda}.
\]
This truncated Hamiltonian acts on a finite-dimensional Hilbert space
and can be simulated on a digital quantum computer or using
tensor-network methods.
For a normalized initial state $\psi_0\in\ran\Pi_{\EE}$, we initialize
the truncated evolution in the normalized projection
$
\psi_{0,\Lambda}
=\frac{P_\Lambda\psi_0}{\|P_\Lambda\psi_0\|},
$
provided that $P_\Lambda\psi_0\neq0$.
Our aim is to choose $\Lambda$ so that this projection is well defined and
\[
\operatorname{err}_{a,\Lambda}(T,\psi_0)
:=\sup_{|t|\leq T}
\left\|
\mathrm e^{-itH_a}\psi_0
-\mathrm e^{-itH_{a,\Lambda}}\psi_{0,\Lambda}
\right\|
\leq\epsilon
\]
for every normalized $\psi_0\in\ran\Pi_{\EE}$. The bound we derive to achieve this is as follows: We first define
\begin{equation}
 F(a,g,T,\epsilon)
 =1+\log\frac{a_{\refscale}}a
       +\log(1+g^{-2})
       +\log\left(1+\frac{T}{T_{\refscale}}\right)
       +\log\frac1\epsilon,
\end{equation}
where $a_{\mathrm{ref}}$ and $T_{\mathrm{ref}}$ are fixed reference lattice spacing and time respectively. Then for general spatial dimension $d$, we obtain the truncation bound, as stated in Theorem~\ref{thm:general},
\begin{equation}
\label{eq:truncate_bound_intro_general}
 \Lambda_{\mathrm c}\leq C_d\left[
       \frac{1}{g}\sqrt{F(a,g,T,\epsilon)+a\EE}
       +F(a,g,T,\epsilon)+a\EE\right].
\end{equation}
For $d=2$, we obtain a tighter bound, as stated in Theorem~\ref{thm:continuum},
\begin{equation}
\label{eq:truncate_bound_intro_2d}
 \Lambda_G\leq
 \frac{C_G}{g}\sqrt{F(a,g,T,\epsilon)+a\EE}.
\end{equation}
Here $C_d$ and $C_G$ are positive constants depending only on the fixed
lattice family and reference scales, and are independent of
$a$, $g$, $\EE$, $T$, and $\epsilon$.

The proof has two key ingredients. The first is a local comparison of
vacuum energies, which bounds the local electric energy
$\frac{h^2}{2}E_\ell^2$ in terms of the \emph{difference} between $K$
and its ground energy $e_0$, up to a correction that vanishes with $h$.
More precisely, Lemma~\ref{lem:local} gives
$\frac{h^2}{2}E_\ell^2\leq K-e_0+\mathcal{O}(h)$.
This allows us to control the electric numbers to be truncated through
the excitation window $\EE$, which can remain fixed as $a\to0$,
without introducing the total vacuum energy.
The $\mathcal{O}(h)$ correction is crucial for obtaining the $1/g$
dependence in the truncation bounds
\eqref{eq:truncate_bound_intro_general} and
\eqref{eq:truncate_bound_intro_2d}.
We obtain this inequality by applying the variational principle to a
suitable trial wavefunction.
Since our focus is on a single link $\ell$, a natural first choice is
the tensor product of a constant wavefunction on $\ell$ and a ground
state of the remaining Hamiltonian.
This choice, however, gives only an $\mathcal{O}(1)$ correction,
leading to $1/g^2$ dependence in the resulting truncation bounds.
Our construction uses an entangled trial wavefunction obtained by
multiplying a reference ground-state wavefunction by a function $f$
that depends jointly on $\ell$ and the remaining links.
Using a ground-state representation, as in
Ref.~\cite{frank2007eigenvalue}, we optimize the parameters of $f$
to obtain the $\mathcal{O}(h)$ correction.

The second ingredient is Agmon's exponentially weighted energy
method~\cite{agmon1982lectures,zworski_book}, which establishes exponential decay of eigenfunctions in classically forbidden regions by controlling their exponentially weighted norms through energy estimates. 
We adapt it to the hopping in the electric
basis as in discrete Agmon estimates~\cite{klein2008agmon}.
We introduce the weight function $\phi_L(n)$ in \eqref{eq:weight}
to penalize large electric numbers exponentially.
Specifically, we define the operator $e^{\phi_L(E_\ell)}$ and bound
the weighted norm $\|e^{\phi_L(E_\ell)}\psi\|$ uniformly over all
normalized $\psi\in\operatorname{ran}(\Pi_{\EE})$.
This yields an electric tail bound with the decay factor
$\exp[-\mathcal S_d(h\Lambda)/(2h)]$, where $\mathcal S_d$ is the
monotonically increasing function constructed in \eqref{eq:action}.
We then convert this tail bound into an evolution error bound
using Duhamel's formula.

While completing this work, we learned of concurrent work Ref.~\cite{liang2026lowenergy}. That work also considers states in
a spectral window above the ground energy and the corresponding
truncated evolution, on a three-dimensional periodic cubic lattice in
the sector with zero global electric flux. Its product tail bound starts
at an electric number asymptotic to $\sqrt8/g^2$ as $g\to0$ at fixed
lattice and excitation window, thus requiring the truncation bound to scale as $1/g^2$. By comparison, our local vacuum comparison and
semiclassical decay estimate allows us to improve this dependence to
$g^{-1}$, up to logarithmic factors.

Table~\ref{tab:intro-cutoffs} compares
sufficient cutoffs in two and three spatial dimensions using the bounds from this work and Refs.~\cite{tong2022provably,liang2026lowenergy}. We use a periodic
hypercube of side length $L_{\mathrm{box}}>0$ with $N\geq3$ sites per
direction, so $a=L_{\mathrm{box}}/N$, $M=dN^d$, and
$N_p=\binom d2N^d$. In both panels,
$T/L_{\mathrm{box}}=1$, $L_{\mathrm{box}}\EE=1$, and
$\epsilon=10^{-3}$.  $\Lambda_{(1)}$ is computed from from
\eqref{eq:full-cutoff} in this work as the smallest integer $\Lambda\geq0$ satisfying
$\mathcal S_d(g^2\Lambda)/(2g^2)\geq L_{T,d}$.
$\Lambda_{(2)}$ is computed using the method described in Ref.~\cite{tong2022provably}, adapted to our setting. The precise way to compute this bound is described in Appendix~\ref{sec:truncation_bound_tong_etal}.
For $d=3$, we also report $\Lambda_{(3)}$ from Ref.~\cite{liang2026lowenergy} as the smallest integer
accepted by the state-norm bound in
Ref.~\cite[Theorem~1, Eqs.~(6)--(9)]{liang2026lowenergy}, restricting the
comparison to the sector with zero global electric flux. Since theorem is
stated only for $d=3$, we do not include a comparison with Ref.~\cite{liang2026lowenergy} for $d=2$.

\begin{table}[!htbp]
 \centering
 \begin{minipage}[t]{0.46\linewidth}
 \centering
 \textbf{(a) $d=2$}\\[0.6em]
 \begin{tabular}{@{}cccc@{}}
  \hline
  $N$ & $g$ & $\Lambda_{(1)}$ (this work)
      & $\Lambda_{(2)}$ \cite{tong2022provably} \\
  \hline
   8 & 0.05 & 230 & $17\,595$ \\
   8 & 0.10 & 112 & $4\,455$ \\
   8 & 0.20 &  55 & $1\,148$ \\
   8 & 0.30 &  37 & $528$ \\
  \hline
  16 & 0.05 & 240 & $34\,997$ \\
  16 & 0.10 & 117 & $8\,808$ \\
  16 & 0.20 &  58 & $2\,238$ \\
  16 & 0.30 &  39 & $1\,014$ \\
  \hline
 \end{tabular}
 \end{minipage}\hfill
 \begin{minipage}[t]{0.52\linewidth}
 \centering
 \textbf{(b) $d=3$}\\[0.6em]
 \begin{tabular}{@{}ccccc@{}}
  \hline
  $N$ & $g$ & $\Lambda_{(1)}$ (this work)
      & $\Lambda_{(2)}$ \cite{tong2022provably} & $\Lambda_{(3)}$ \cite{liang2026lowenergy} \\
  \hline
   8 & 0.05 & 297 & $35\,047$ & $1\,296$ \\
   8 & 0.10 & 146 & $8\,833$ & 364 \\
   8 & 0.20 &  72 & $2\,251$ & 111 \\
   8 & 0.30 &  48 & $1\,023$ & 58 \\
  \hline
  16 & 0.05 & 312 & $69\,849$ & $1\,305$ \\
  16 & 0.10 & 153 & $17\,537$ & 369 \\
  16 & 0.20 &  76 & $4\,430$ & 114 \\
  16 & 0.30 &  51 & $1\,993$ & 60 \\
  \hline
 \end{tabular}
 \end{minipage}
 \caption{Computed sufficient electric cutoffs for (a) $d=2$ and
 (b) $d=3$, with $T/L_{\mathrm{box}}=L_{\mathrm{box}}\EE=1$ and
 $\epsilon=10^{-3}$. $\Lambda_{(1)-(3)}$ are computed from this work, Ref.~\cite{tong2022provably}, and Ref.~\cite{liang2026lowenergy} respectively.
 All columns guarantee state-norm error at most $\epsilon$ uniformly over
 the spectral window; the $d=3$ comparison is restricted to the sector
 with zero global electric flux to satisfy the requirement in Ref.~\cite{liang2026lowenergy} (no other works require this).}
 \label{tab:intro-cutoffs}
\end{table}

\section{Problem Setup}\label{sec:setup}

\subsection{Lattice and physical Hilbert space}

\begin{definition}[Lattice family]\label{def:lattice}
We fix the spatial dimension to be $d\geq 2$, and consider a $d$-dimensional spatial region in the form of a box with side lengths
$L_1,\ldots,L_d>0$ and volume $\mathcal V=\prod_{i=1}^dL_i$. 
It has either open or periodic boundary conditions in all directions. We also fix a
reference lattice spacing $a_{\refscale}>0$. A spacing $0<a\leq a_{\refscale}$
is \emph{admissible} if $N_i=L_i/a$ is a positive integer for every
$i\in\{1,\ldots,d\}$ (in the periodic case we require $N_i\geq3$).
For each admissible spacing $a$, we partition the box into hypercubic
cells of side length $a$, with $N_i=L_i/a$ cells along direction $i$.
This gives us a lattice formed by the resulting sets of vertices, oriented links, and elementary square
plaquettes, which we denote by $\mathsf V$, $\mathsf L$, and $\mathsf P$, respectively.
We write $M=|\mathsf L|$ and $N_p=|\mathsf P|$.
Each plaquette has four distinct links. Denoting by $r_\ell$ the number of
plaquettes incident to a link $\ell$, we have 
\begin{equation}\label{eq:coordination}
 r_\ell\leq R_d:=2(d-1).
\end{equation}
\end{definition}

Throughout this paper we consider $d$ as a fixed number, and other constants may depend on $d$.
The restriction $d\geq2$ is included to ensure that spatial plaquettes are present.

Links and plaquettes are related through the plaquette--link
incidence matrix $C\in\mathbb Z^{N_p\times M}$.
We first choose one orientation for each link and plaquette. The entry
$C_{p\ell}$ is $+1$ if the orientation of $\ell$ agrees with the
oriented boundary of $p$, $-1$ if it is opposite, and zero if
$\ell$ does not belong to that boundary.
For $\theta=(\theta_\ell)_{\ell\in\mathsf L}\in\TT^M$, where
$\TT=\mathbb R/(2\pi\mathbb Z)$, the plaquette angle is
$\theta_p=(C\theta)_p\pmod{2\pi}$. 

We will use the following normalized Lebesgue measure
\[
 d\nu_m(\theta)=\prod_{j=1}^m\frac{d\theta_j}{2\pi}.
\]
The full link Hilbert space is $\HH_{\mathrm{link}}=L^2(\TT^M,d\nu_M)$.
Its orthonormal electric basis consists of
$e_n(\theta)=\mathrm e^{in\cdot\theta}$, $n\in\mathbb Z^M$, and the electric
operator $E_\ell=-i\partial_{\theta_\ell}$ satisfies
$E_\ell e_n=n_\ell e_n$.
We use $\|\cdot\|$ for the Hilbert-space norm and the induced operator
norm, and $\one_I(A)$ for the spectral projection of a self-adjoint
operator $A$ onto a Borel set $I\subseteq\mathbb R$.

Similar to the plaquette--link incidence matrix, we also define the vertex--link incidence
matrix $D\in\mathbb Z^{|\mathsf V|\times M}$, with $D_{v\ell}=1$ at the head of $\ell$, $-1$ at its tail, and zero elsewhere. Because a plaquette
boundary has no endpoints, we have $DC^{\mathsf T}=0$.
Figure~\ref{fig:incidence-plaquettes} illustrates these conventions
on two adjacent plaquettes, including the opposite signs of their
shared link in the two rows of $C$.

\begin{figure}[!tbp]
 \centering
 \small
 \begin{minipage}[t]{0.5\linewidth}
  \centering
  \textbf{(a) Two oriented plaquettes}\\[0.6em]
  \begin{tikzpicture}[
    x=1cm,y=1cm,
    link/.style={-{Latex[length=2mm]},line width=0.9pt,
                 shorten <=3pt,shorten >=3pt},
    boundary/.style={-{Latex[length=1.8mm]},dashed,
                     orange!75!black,line width=0.8pt},
    vertex/.style={circle,fill=black,inner sep=1.7pt},
    every node/.style={font=\small}
  ]
   \coordinate (v1) at (0,0);
   \coordinate (v2) at (2.4,0);
   \coordinate (v3) at (2.4,2.4);
   \coordinate (v4) at (0,2.4);
   \coordinate (v5) at (4.8,0);
   \coordinate (v6) at (4.8,2.4);
   \fill[black!3] (v1) rectangle (v3);
   \fill[black!3] (v2) rectangle (v6);
   \draw[link] (v1) -- node[below=5pt] {$\ell_1$} (v2);
   \draw[link,blue!65!black] (v2) --
       node[fill=white,inner sep=2pt] {$\ell_2$} (v3);
   \draw[link] (v4) -- node[above=5pt] {$\ell_3$} (v3);
   \draw[link] (v1) -- node[left=5pt] {$\ell_4$} (v4);
   \draw[link] (v2) -- node[below=5pt] {$\ell_5$} (v5);
   \draw[link] (v5) -- node[right=5pt] {$\ell_6$} (v6);
   \draw[link] (v3) -- node[above=5pt] {$\ell_7$} (v6);
   \foreach \offset in {0,2.4} {
    \begin{scope}[xshift=\offset cm]
     \draw[boundary] (0.45,0.45) -- (1.95,0.45);
     \draw[boundary] (1.95,0.45) -- (1.95,1.95);
     \draw[boundary] (1.95,1.95) -- (0.45,1.95);
     \draw[boundary] (0.45,1.95) -- (0.45,0.45);
    \end{scope}
   }
   \node at (1.05,1.2) {$p_1$};
   \node at (3.75,1.2) {$p_2$};
   \node[orange!75!black] at (1.65,1.2) {$+1$};
   \node[orange!75!black] at (3.15,1.2) {$-1$};
   \node[vertex,label=below left:$v_1$] at (v1) {};
   \node[vertex,label=below:$v_2$] at (v2) {};
   \node[vertex,label=above:$v_3$] at (v3) {};
   \node[vertex,label=above left:$v_4$] at (v4) {};
   \node[vertex,label=below right:$v_5$] at (v5) {};
   \node[vertex,label=above right:$v_6$] at (v6) {};
   \node[blue!65!black] at (2.4,-0.7) {tail: $-1$};
   \node[blue!65!black] at (2.4,3.1) {head: $+1$};
  \end{tikzpicture}
 \end{minipage}\hfill
 \begin{minipage}[t]{0.49\linewidth}
  \centering
  \textbf{(b) The two incidence matrices}
  \setlength{\arraycolsep}{3.3pt}
  \[
   C_{\mathrm{loc}}=
   \begin{array}{c|rrrrrrr}
        &\ell_1&\ell_2&\ell_3&\ell_4&\ell_5&\ell_6&\ell_7\\ \hline
    p_1 &+1&\textcolor{orange!75!black}{+1}&-1&-1&0&0&0\\
    p_2 &0&\textcolor{orange!75!black}{-1}&0&0&+1&+1&-1
   \end{array}
  \]
  \[
   D_{\mathrm{loc}}=
   \begin{array}{c|rrrrrrr}
        &\ell_1&\ell_2&\ell_3&\ell_4&\ell_5&\ell_6&\ell_7\\ \hline
    v_1 &-1&0&0&-1&0&0&0\\
    v_2 &+1&\textcolor{blue!65!black}{-1}&0&0&-1&0&0\\
    v_3 &0&\textcolor{blue!65!black}{+1}&+1&0&0&0&-1\\
    v_4 &0&0&-1&+1&0&0&0\\
    v_5 &0&0&0&0&+1&-1&0\\
    v_6 &0&0&0&0&0&+1&+1
   \end{array}
  \]
 \end{minipage}

 \caption{Incidence matrices for two adjacent plaquettes.
 Solid arrows give the link orientations; dashed arrows traverse
 both plaquette boundaries counterclockwise. Each row of $C$ sums
 link angles around one plaquette. The shared link $\ell_2$
 contributes $+1$ to row $p_1$ and $-1$ to row $p_2$
 (orange); links absent from a boundary contribute zero.
 Each column of $D$ has $-1$ at the link's tail and $+1$ at its
 head, illustrated in blue for $\ell_2$.
 Thus $D^{\mathsf T}$ takes vertex-angle differences along links.
 Each oriented plaquette boundary has zero net incidence at every
 vertex, giving $D_{\mathrm{loc}}C_{\mathrm{loc}}^{\mathsf T}
 =0_{6\times2}$ and, for the full lattice, $DC^{\mathsf T}=0$.}
 \label{fig:incidence-plaquettes}
\end{figure}

A gauge transformation acts by
$\theta\mapsto\theta+D^{\mathsf T}\alpha$, with
$\alpha\in\TT^{|\mathsf V|}$. We work with pure compact U(1) gauge theory
with zero background charge and no matter or external charges. Its
physical Hilbert space is
\begin{equation*}
 \HH_{\phys}
 =\{\psi\in\HH_{\mathrm{link}}:
       \psi(\theta+D^{\mathsf T}\alpha)=\psi(\theta)
       \text{ for every }\alpha\}
 =\overline{\operatorname{span}}\{e_n:Dn=0\}.
\end{equation*}
These constraints $\psi(\theta+D^{\mathsf T}\alpha)=\psi(\theta)$ are called \emph{Gauss's law}.

\subsection{Hamiltonian and low-energy states}

For independent parameters $a>0$ and dimensionless coupling $g>0$, we define the Kogut--Susskind Hamiltonian for the U(1) pure gauge theory:
\begin{equation}\label{eq:hamiltonian}
 H_a=\frac1a\left[
       \frac{g^2}{2}\sum_{\ell\in\mathsf L}E_\ell^2
       +\frac1{g^2}\sum_{p\in\mathsf P}(1-\cos\theta_p)\right].
\end{equation}
We leave the dependence on $g$ implicit in $H_a$ and the quantities defined below.
We first regard this as a self-adjoint operator on $\HH_{\mathrm{link}}$,
with domain $H^2(\TT^M)$ and quadratic-form domain $H^1(\TT^M)$, where
$H^k$ denotes the periodic Sobolev space with square-integrable weak
derivatives through order $k$. We then restrict it to the invariant
subspace $\HH_{\phys}$. 
Our aim is to improve the cutoff bound at small $g$. To use tools from semiclassical analysis, we rescale the Hamiltonian as follows:
\begin{equation}\label{eq:rescaling}
 K=ahH_a=-\frac{h^2}{2}\Delta_\theta+V(\theta),
 \quad V(\theta)=\sum_{p\in\mathsf P}[1-\cos(C\theta)_p],\quad h=g^2,
\end{equation}
where $\Delta_\theta=\sum_{\ell\in\mathsf L}\partial_{\theta_\ell}^2$ is the Laplacian operator.
This rescaling puts $K$ in semiclassical Schr\"odinger form:
$h$ is the dimensionless semiclassical parameter, and weak lattice
coupling corresponds to $h\to0$. Following the notation in \cite{zache2022toward}, we use
$e_{\mathrm{charge}}>0$ to denote the bare electric charge in $d$ spatial
dimensions. One case of interest is refinement at fixed bare charge, where
\begin{equation}\label{eq:continuum-coupling}
 h=g^2=e_{\mathrm{charge}}^2a^{3-d}.
\end{equation}
Thus $e_{\mathrm{charge}}^2$ has dimensions
$\text{length}^{\,d-3}$. 
With $e_{\mathrm{charge}}$ fixed, \eqref{eq:continuum-coupling} makes $h$ tend to zero
when $d=2$, keeps it constant when $d=3$, and makes it grow when
$d>3$. This affects the bound we obtain for different spatial dimensions.
We also consider other choices $g=g(a)$, corresponding to a bare charge
that varies with $a$. Our general bounds do not require fixed bare charge.

Let $E_0(a)=\inf\sigma(H_a|_{\HH_{\phys}})$ and $e_0=ahE_0(a)$.
These are the absolute interacting vacuum energies. $E_0(a)$ is also equal to the
full-space ground energy, i.e., 
$E_0(a)=\inf\sigma(H_a)$, and similarly for $e_0$.
This is because the elliptic operator $K$ on the connected
compact torus has a unique normalized strictly positive ground state wavefunction.
Gauge transformations commute with $K$ and preserve positivity and
normalization, so uniqueness makes this wavefunction gauge invariant.

\begin{definition}[Excitation window]\label{def:window}
For a finite physical excitation window $\EE\geq0$, define on
$\HH_{\phys}$
\begin{equation}\label{eq:window}
 \Pi_{\EE}=\one_{[E_0(a),E_0(a)+\EE]}(H_a)
          =\one_{[e_0,e_0+\delta]}(K),
 \quad \delta=ah\EE.
\end{equation}
The allowed initial states are all normalized vectors in
$\ran\Pi_{\EE}$, including arbitrary superpositions.
\end{definition}

Unless stated otherwise, $\EE$ may depend on $a$; the cutoff bounds
below retain this dependence explicitly.

At a fixed finite lattice, the kinetic operator
$-(h^2/2)\Delta_\theta$ has compact resolvent, since its Fourier
eigenvalues $h^2|n|^2/2$ tend to infinity as $|n|\to\infty$.
The potential $V$ is bounded, so $K$ also has compact resolvent
\cite[Theorem~IV.3.17]{kato1995perturbation}.
Consequently, the spectral subspace $\ran\Pi_{\EE}$ is finite-dimensional
\cite[Theorem~III.6.29]{kato1995perturbation}.
We do not assume any bound on its dimension or spectral gap.

\subsection{Electric truncation and error criterion}

\begin{definition}[Truncated evolution]\label{def:truncation}
For an integer $\Lambda\geq0$, define on $\HH_{\phys}$
\begin{equation*}
 P_\Lambda=\prod_{\ell\in\mathsf L}\one_{|E_\ell|\leq\Lambda},
 \quad
 H_{a,\Lambda}
 =P_\Lambda H_aP_\Lambda\big|_{\ran P_\Lambda}.
\end{equation*}
If $P_\Lambda\psi_0\neq0$, set
$\psi_{0,\Lambda}=P_\Lambda\psi_0/\|P_\Lambda\psi_0\|$.
For $T\geq0$, the truncation error is
\begin{equation}\label{eq:error}
 \operatorname{err}_{a,\Lambda}(T,\psi_0)
 =\sup_{|t|\leq T}
   \left\|\mathrm e^{-itH_a}\psi_0
     -\mathrm e^{-itH_{a,\Lambda}}\psi_{0,\Lambda}\right\|.
\end{equation}
We identify $\ran P_\Lambda$ with its subspace in $\HH_{\phys}$ when
comparing the two evolutions.
\end{definition}

All factors of $P_\Lambda$ are diagonal in the electric basis, so the
cutoff preserves Gauss's law. The compressed Hamiltonian is finite
dimensional, with local Hilbert space dimension $2\Lambda+1$ on each link. Our task is to
choose $\Lambda$ so that \eqref{eq:error} is below a prescribed precision target
$0<\epsilon<1$, uniformly over the states in Definition~\ref{def:window}.

\section{Main Results}\label{sec:results}

We first give a cutoff $\Lambda$ that ensures truncation error at most
$\epsilon$ for every fixed $d\geq2$ and every simulation time $T\geq0$.
The cutoff depends on $T$, so this result places no upper restriction on
the simulation time. We then state the truncation error bound from which
this cutoff follows and the tighter bound in two spatial dimensions.

For a finite lattice, $h>0$, a finite excitation window $\EE\geq0$,
$T\geq0$, and $0<\epsilon<1$, we define
\begin{equation}\label{eq:target}
 \begin{aligned}
 &A_d(a)=\frac{2(R_d+a\EE)}{\sqrt{R_d}}, \quad
 \gamma_d=4\sqrt{2R_d}=8\sqrt{d-1},\\
 &\mathcal B_T=2\sqrt M+\frac{2N_pT}{ah},
 \quad L_{T,d}=A_d(a)+\log\frac{\sqrt2\mathcal B_T}{\epsilon}.
 \end{aligned}
\end{equation}
Here $A_d(a)$ controls the weighted mass near small electric fields,
and $L_{T,d}>0$ is the logarithmic target for the tail exponent.
Their dependence on $\EE$ is through the term $2a\EE/\sqrt{R_d}$.
For a fixed reference time $T_{\refscale}>0$, also set
\begin{equation}\label{eq:F}
 F(a,g,T,\epsilon)
 =1+\log\frac{a_{\refscale}}a
       +\log(1+g^{-2})
       +\log\left(1+\frac{T}{T_{\refscale}}\right)
       +\log\frac1\epsilon.
\end{equation}
All logarithms have dimensionless arguments, and $F\geq1$.

\begin{theorem}[Main result: cutoff bound for target accuracy]\label{thm:general}
We fix the lattice family
in Definition~\ref{def:lattice} with reference lattice spacing $a_{\refscale}$
and a reference time $T_{\refscale}>0$.
For each admissible $a$ and each $g>0$, set $h=g^2$ and consider the Hamiltonian
\eqref{eq:hamiltonian} and the truncated Hamiltonian $H_{a,\Lambda}$ in
Definition~\ref{def:truncation}. For any finite $\EE\geq0$, $T\geq0$,
and $0<\epsilon<1$, choose
\begin{equation}\label{eq:coarse-cutoff}
 \Lambda_{\mathrm c}
 =\left\lceil\max\left\{
       \frac{1}{g}\sqrt{\gamma_d L_{T,d}},\,
       \gamma_d L_{T,d}\right\}\right\rceil.
\end{equation}
Then every normalized $\psi_0\in\ran\Pi_{\EE}$ has
$P_{\Lambda_{\mathrm c}}\psi_0\neq0$, and for the error defined in \eqref{eq:error},
\begin{equation}\label{eq:general-error}
 \operatorname{err}_{a,\Lambda_{\mathrm c}}(T,\psi_0)\leq\epsilon.
\end{equation}
There is a constant $C_d>0$, depending only on the fixed lattice family
and reference time, such that, with $F$ as in \eqref{eq:F}, for all
admissible $(a,g,\EE,T,\epsilon)$,
\begin{equation}\label{eq:general-scaling}
 \Lambda_{\mathrm c}\leq C_d\left[
       \frac{1}{g}\sqrt{F(a,g,T,\epsilon)+a\EE}
       +F(a,g,T,\epsilon)+a\EE\right].
\end{equation}
The constant is independent of $a,g,\EE,T,\epsilon$, the initial state,
and the rank of $\Pi_{\EE}$. No bound uniform in $d$ is asserted.
\end{theorem}

For the fixed-bare-charge family \eqref{eq:continuum-coupling},
the coupling term satisfies
$\log(1+g^{-2})=\mathcal{O}(1+\log(a_{\refscale}/a))$,
with a constant that may depend on
$e_{\mathrm{charge}}$. Substituting $g=e_{\mathrm{charge}}a^{(3-d)/2}$
in \eqref{eq:general-scaling} gives the corresponding refinement bound.

This cutoff bound is a consequence of the truncation error bound that we are going to introduce below.
We introduce the function $\mathcal S_d:\mathbb R\to[0,\infty)$:
\begin{equation}\label{eq:action}
 \begin{aligned}
 \mathcal S_d(p)
 =\int_0^{|p|}\arcosh\left(1+\frac{s^2}{2R_d}\right)\,ds
 =2|p|\arsinh\frac{|p|}{2\sqrt{R_d}}
       -2\sqrt{p^2+4R_d}+4\sqrt{R_d}.
 \end{aligned}
\end{equation}
This function is continuous, strictly increasing on $[0,\infty)$, and unbounded.
We write $\mathcal S_d^{-1}$ for its inverse on that interval.
The function $\mathcal S_d$ determines the exponential weight used to
bound large electric numbers. In the electric representation, each
incident plaquette shifts the electric number on a link by $\pm1$.
Pairing these shifts after exponential weighting introduces a hyperbolic
cosine, as in the discrete Agmon estimates of
Klein and Rosenberger~\cite[Lemma~3.1]{klein2008agmon}.
For a link electric number $n\in\mathbb Z$, write $p=hn$.
The corresponding electric energy in $K$ is $p^2/2$.
We choose $\mathcal S_d$ so that its derivative satisfies
\begin{equation}\label{eq:action-balance}
 R_d\bigl[\cosh\mathcal S_d'(p)-1\bigr]=\frac{p^2}{2},
 \quad p\geq0.
\end{equation}
This balances the increase caused by the weight against the electric
energy $p^2/2$. Solving for $\mathcal S_d'(p)$ and integrating from zero gives \eqref{eq:action}. 

\begin{theorem}[Truncation error bound]\label{thm:full}
Fix a finite lattice satisfying Definition~\ref{def:lattice}, and
let $a>0$ and $h=g^2>0$. 
For $\EE\geq0$ and an integer $\Lambda\geq0$, set
\[
 q_\Lambda=\sqrt2\exp\left[
           A_d(a)-\frac{\mathcal S_d(h\Lambda)}{2h}\right].
\]
For every normalized $\psi_0\in\ran\Pi_{\EE}$ with
$P_\Lambda\psi_0\neq0$ and every $T\geq0$,
\begin{equation}\label{eq:full-error}
 \operatorname{err}_{a,\Lambda}(T,\psi_0)
 \leq\mathcal B_Tq_\Lambda
 =\sqrt2\mathcal B_T
      \exp\left[A_d(a)-\frac{\mathcal S_d(h\Lambda)}{2h}\right].
\end{equation}
The projected state is nonzero whenever $\sqrt Mq_\Lambda<1$.
\end{theorem}
This theorem applies to arbitrary $g>0$, without requiring fixed bare charge.
$L_{T,d}$ in \eqref{eq:target} is chosen so that
\eqref{eq:full-error} is at most $\epsilon$ whenever
\begin{equation}\label{eq:action-target}
 \frac{\mathcal S_d(h\Lambda)}{2h}\geq L_{T,d}.
\end{equation}
A sufficient integer cutoff satisfying \eqref{eq:action-target}
at all parameters is therefore
\begin{equation}\label{eq:full-cutoff}
 \Lambda_{\mathrm{full}}
 =\left\lceil\frac{\mathcal S_d^{-1}(2hL_{T,d})}h\right\rceil.
\end{equation}
For this choice of cutoff, $\sqrt M q_\Lambda\leq \mathcal{B}_T q_\Lambda/2\leq\epsilon/2<1$,
so $P_\Lambda\psi_0\neq0$ and the normalized projection is well defined.
The explicit cutoff \eqref{eq:coarse-cutoff}
is obtained from the following lower bound for the function $\mathcal{S}_d(p)$ in \eqref{eq:action}:
\begin{equation}\label{eq:global-action}
 \mathcal S_d(p)\geq\frac{2}{\gamma_d}\min\{p^2,p\},
 \quad p\geq0.
\end{equation}
We prove this bound in Appendix~\ref{app:action-bounds}.

A tighter characterization of $\mathcal{S}_d(p)$ can lead to an improved cutoff bound in certain parameter regimes, as stated below:
\begin{corollary}[Improved cutoff bound in the Gaussian regime]\label{cor:gaussian}
Under the hypotheses of Theorem~\ref{thm:full}, define
\begin{equation}\label{eq:general-gaussian}
 \Lambda_{G,d}
 =\left\lceil\sqrt{\frac{\gamma_d L_{T,d}}h}\right\rceil.
\end{equation}
If $h\Lambda_{G,d}\leq2\sqrt{R_d}$, then
$P_{\Lambda_{G,d}}\psi_0\neq0$ and
$\operatorname{err}_{a,\Lambda_{G,d}}(T,\psi_0)\leq\epsilon$
for every normalized $\psi_0\in\ran\Pi_{\EE}$.
A sufficient condition is
\begin{equation}\label{eq:general-rounding}
 \sqrt{\gamma_d hL_{T,d}}+h\leq2\sqrt{R_d}.
\end{equation}
\end{corollary}
Since $h=g^2$, this cutoff is $\lceil\sqrt{\gamma_d L_{T,d}}/g\rceil$,
with further dependence on $g$ only logarithmic through $L_{T,d}$.
We say this cutoff bound works in the \emph{Gaussian regime} because when $h\Lambda_{G,d}\leq 2\sqrt{R_d}$ we will lower bound $\mathcal{S}_d(p)$ using a quadratic function, resulting in a Gaussian-type decay of the truncation error.
This is especially relevant for the $d=2$ case,
because fixed bare charge makes $h$ tend to zero as the lattice spacing $a$
decreases, ensuring $h\Lambda_{G,d}\leq2\sqrt{R_d}$ will hold for sufficiently small $a$, when $\EE$ is fixed and $T$ and $1/\epsilon$ are polynomials of $1/a$.

\begin{theorem}[Improved cutoff bound for $d=2$]\label{thm:continuum}
Under the conditions of
Theorem~\ref{thm:general}, we further assume $d=2$. Then
$R_2=2$, $\gamma_2=8$, $h=g^2$, and
\[
 L_{T,2}=\sqrt2(2+a\EE)+\log\frac{\sqrt2\mathcal B_T}{\epsilon}.
\]
For an admissible $a$, any $g>0$, finite $\EE\geq0$, $T\geq0$, and
$0<\epsilon<1$, set
\begin{equation}\label{eq:gaussian-cutoff}
 \Lambda_G=\left\lceil\frac{1}{g}\sqrt{8L_{T,2}}\right\rceil.
\end{equation}
We assume for this cutoff,
\begin{equation}\label{eq:gaussian-condition}
 g^2\Lambda_G\leq2\sqrt2.
\end{equation}
Then every normalized $\psi_0\in\ran\Pi_{\EE}$ has
$P_{\Lambda_G}\psi_0\neq0$, and
\begin{equation}\label{eq:main-error}
 \operatorname{err}_{a,\Lambda_G}(T,\psi_0)\leq\epsilon.
\end{equation}
There is a constant $C_G>0$, depending only on the fixed spatial region,
boundary conditions, $a_{\refscale}$, and $T_{\refscale}$, such that
\begin{equation}\label{eq:main-scaling}
 \Lambda_G\leq
 \frac{C_G}{g}\sqrt{F(a,g,T,\epsilon)+a\EE}.
\end{equation}
This bound is uniform over all admissible $(a,g,\EE,T,\epsilon)$ satisfying
\eqref{eq:gaussian-condition}, and over all states in the spectral
window, independently of its rank.
\end{theorem}

In this case \eqref{eq:general-rounding} becomes
\begin{equation}\label{eq:rounding-condition}
 g\sqrt{8L_{T,2}}+g^2\leq2\sqrt2.
\end{equation}
Theorem~\ref{thm:continuum} bounds the cutoff by a constant times
$g^{-1}\sqrt{F(a,g,T,\epsilon)+a\EE}$, so the excitation-window term
also enters under a square root.
It only works in the parameter regime where \eqref{eq:gaussian-condition} holds.
When this condition fails, Theorems~\ref{thm:general}
and~\ref{thm:full} continue to apply.
At fixed bare charge, Theorem~\ref{thm:continuum} leads to the following improved refinement bound for $d=2$:

\begin{corollary}[Polynomial times and inverse errors for $d=2$]
\label{cor:polynomial}
Under the fixed parameters of Theorem~\ref{thm:continuum}, also fix a
bare charge $e_{\mathrm{charge}}>0$ and impose $g^2=e_{\mathrm{charge}}^2a$.
Fix a finite excitation window $\EE\geq0$, $\kappa,\eta\geq0$, and
$0<\epsilon_{\refscale}<1$, and set
\[
 T=T_{\refscale}(a_{\refscale}/a)^\kappa,\quad
 \epsilon=\epsilon_{\refscale}(a/a_{\refscale})^\eta.
\]
For all sufficiently small admissible $a$,
\eqref{eq:gaussian-condition} holds and the error guarantee
\eqref{eq:main-error} applies. Along any sequence of admissible
spacings tending to zero,
\begin{equation}\label{eq:polynomial-asymptotic}
 \begin{aligned}
 L_{T,2}&=(4+\kappa+\eta)\log(a_{\refscale}/a)+\mathcal{O}(1),\\
 \Lambda_G&\sim
 \sqrt{\frac{8(4+\kappa+\eta)}{e_{\mathrm{charge}}^2a}
                    \log\frac{a_{\refscale}}a}.
 \end{aligned}
\end{equation}
\end{corollary}

Thus this improved bound holds with polynomial growth of time and inverse error, giving an almost (up to a polylogarithmic factor) $a^{-1/2}$-dependence in two spatial dimensions.

\section{Proof of the Main Results}\label{sec:proof}

We first prove the truncation error bound in Theorem~\ref{thm:full},
then derive the cutoff bounds from it. The main step is to control the
probability of a large electric number on one link uniformly over the
spectral window. A direct bound using the total energy would introduce
the vacuum energy of the whole lattice. To obtain a bound that depends
only on the terms incident to the chosen link, we compare $K$ with a
Hamiltonian in which the electric
term on one link and its incident magnetic terms have been removed.
This comparison gives a local electric-energy bound. We then use an
exponential weight to turn that bound into an electric tail estimate,
and apply Duhamel's formula to control the evolution error.

\subsection{A local comparison of vacuum energies}\label{sec:local}

We fix a link $\ell$ and let
$S=\{p\in\mathsf P:C_{p\ell}\neq0\}$ be the set of plaquettes incident to this link, so $|S|=r_\ell$.
To compare the vacuum energies before and after removing the terms
incident to this link, we define
\begin{equation}\label{eq:removed}
 R_\ell=\frac{h^2}{2}\sum_{j\neq\ell}E_j^2
      +\sum_{p\notin S}[1-\cos(C\theta)_p],
 \quad e_R=\inf\sigma(R_\ell).
\end{equation}
The operator $R_\ell$ acts trivially on the omitted link:
$R_\ell=I_\ell\otimes\overline R_\ell$, where $I_\ell$ is the identity
on the omitted link space $L^2(\TT)$ and $\overline R_\ell$
acts on the remaining $M-1$ link angles. The operator
$\overline R_\ell$ has a unique positive normalized ground state
wavefunction. We extend this function to a ground state wavefunction of $R_\ell$ by setting the wavefunction to be constant in $\theta_\ell$.

We will multiply this ground state by a function of the removed
plaquette angles $\theta_p$ for $p\in S$ to construct a trial state for $K$. The energy of this trial state then upper bounds the ground state energy $e_0$. The following two
lemmas provide the tools to bound the energy of this trial state.

\begin{lemma}\label{lem:electric-positive}
There is a normalized ground state $\psi_R$ of $R_\ell$, constant in
$\theta_\ell$, with the $L^2$ Fourier expansion
\begin{equation*}
 \psi_R(\theta)=\sum_{n\in\mathbb Z^M}c_n\mathrm e^{in\cdot\theta},
 \quad c_n\geq0,\quad \sum_n c_n^2=1.
\end{equation*}
Moreover, $c_n=0$ whenever $n_\ell\neq0$.
\end{lemma}

The proof is stated in Appendix~\ref{app:semigroup}. Here we will outline the high-level idea.
We evolve the constant function (i.e., it is $1$ for every $\theta$) in imaginary time
under $\overline R_\ell$. The evolution preserves nonnegative Fourier
coefficients in the electric basis. After normalization, it converges
to the ground state wavefunction, whose coefficients are therefore also
nonnegative.

Let $C_S\in\mathbb Z^{r_\ell\times M}$ be the restriction of $C$ to
the rows in $S$. We define the probability measure $\mu$ on
$\TT^{r_\ell}$ by
\begin{equation}\label{eq:mu}
 \int_{\TT^{r_\ell}}F(x)\,d\mu(x)
 =\int_{\TT^M}F(C_S\theta)|\psi_R(\theta)|^2\,d\nu_M(\theta)
\end{equation}
for bounded measurable $F$. In other words, $\mu$ is the distribution of
$x=C_S\theta=(C\theta)_S\pmod{2\pi}$ in the state $\psi_R$.

\begin{lemma}[Nonnegative plaquette Fourier coefficients]\label{lem:mu-positive}
For every $k\in\mathbb Z^{r_\ell}$,
\begin{equation*}
 \widehat\mu(k)
 :=\int_{\TT^{r_\ell}}\mathrm e^{-ik\cdot x}\,d\mu(x)
 =\sum_{n\in\mathbb Z^M}c_nc_{n+C_S^{\mathsf T}k}\geq0.
\end{equation*}
\end{lemma}
\begin{proof}
Put $b=C_S^{\mathsf T}k\in\mathbb Z^M$. By \eqref{eq:mu} and
Parseval's identity,
\[
 \widehat\mu(k)
 =\langle\psi_R,\mathrm e^{-ib\cdot\theta}\psi_R\rangle
 =\sum_n\overline{c_n}c_{n+b}.
\]
Cauchy--Schwarz gives
$\sum_n|c_nc_{n+b}|\leq(\sum_n c_n^2)^{1/2}
(\sum_n c_{n+b}^2)^{1/2}=1$, so the sum is absolutely convergent.
Every summand is nonnegative by Lemma~\ref{lem:electric-positive}.
\end{proof}

These two lemmas help us establish the energy comparison we are aiming for, as stated below:

\begin{lemma}[Local vacuum-energy inequality]\label{lem:local}
For every link $\ell$, the Hamiltonians $K$ in \eqref{eq:rescaling} and $R_\ell$ in
\eqref{eq:removed} satisfy
\begin{equation}\label{eq:vacuum-comparison}
 0\leq e_0-e_R\leq r_\ell h.
\end{equation}
Consequently,
\begin{equation}\label{eq:local}
 K-e_0\geq\frac{h^2}{2}E_\ell^2-r_\ell h.
\end{equation}
This inequality holds in the sense of quadratic forms on
$\HH_{\mathrm{link}}$, and hence on $\HH_{\phys}$, for the lattices in Definition~\ref{def:lattice}.
\end{lemma}
\begin{proof}
All removed magnetic terms are nonnegative, so the lower bound $0\leq e_0-e_R$ in \eqref{eq:vacuum-comparison} holds, and
\begin{equation}\label{eq:local-reduction}
 K\geq R_\ell+\frac{h^2}{2}E_\ell^2
       \geq e_R+\frac{h^2}{2}E_\ell^2.
\end{equation}
It therefore suffices to prove $e_0\leq e_R+r_\ell h$.
We will prove this by constructing a trial state whose energy is upper bounded by $e_R+r_\ell h$.
This is done by multiplying $\psi_R$ by a periodic
function of the removed plaquette angles that concentrates them
near zero, thereby making the magnetic energy on those plaquettes small. At the same time, too much concentration will result in large electric energy, and an appropriate parameter choice is necessary to balance these two effects.
For $s>0$, let
\[
 f_s(x)=\left(\sum_{z\in\mathbb Z^{r_\ell}}
           \mathrm e^{-|x+2\pi z|^2/(2s)}\right)^{1/2},
 \quad x\in\TT^{r_\ell},
\]
the square root of a Gaussian summed over all $2\pi$ translates.
We use the normalized state 
\[
 \Psi_s(\theta)=
 \frac{f_s(C_S\theta)\psi_R(\theta)}
      {\|f_s(C_S\cdot)\psi_R\|}
\]
as the trial state.
The parameter $s$ controls the spread of the trial factor. We choose
it to balance the electric cost of localization against the magnetic
energy. In Appendix~\ref{app:trial},
we use Lemma~\ref{lem:mu-positive} to bound the required moments and
the equation $R_\ell\psi_R=e_R\psi_R$ to obtain
\begin{equation}\label{eq:trial-cost}
 \langle\Psi_s,K\Psi_s\rangle-e_R
 \leq\frac{r_\ell h^2}{2s}+\frac{r_\ell s}{2}.
\end{equation}
Taking $s=h$ and applying the variational principle proves the upper
bound in \eqref{eq:vacuum-comparison}. Combining this with \eqref{eq:local-reduction} we also have \eqref{eq:local}.
\end{proof}

The significance of \eqref{eq:local} is that the correction is both
independent of the number of links and proportional to $h$.
Using $\psi_R$ itself as a trial state, with the wavefunction on link
$\ell$ constant and independent of the remaining links, gives
$\langle\psi_R,K\psi_R\rangle=e_R+r_\ell$. This direct
product construction therefore gives a correction that does not
vanish as $h\to0$. To obtain the correction $r_\ell h$, which is
smaller when $h<1$,
we use the factor $f_s(C_S\theta)$, which depends jointly on
$\theta_\ell$ and the other link angles in the incident plaquettes.
Choosing $s=h$ balances the kinetic cost of this factor against
the restored magnetic energy, as quantified in \eqref{eq:trial-cost}.
The resulting $\mathcal{O}(h)$ correction allows us to control the electric
energy on one link relative to the excitation energy $K-e_0$,
independently of the total vacuum energy, and is essential to the
subsequent Gaussian cutoff bound in Corollary~\ref{cor:gaussian}, including
the fixed-bare-charge case for $d=2$ in Corollary~\ref{cor:polynomial}.

\subsection{An electric tail bound}
\label{sec:tail}

We now turn \eqref{eq:local} into a bound on the probability of large
electric numbers. The method is to bound an exponentially weighted norm:
a bound on this norm forces the unweighted state to have small tails through Markov's inequality.
For the fixed link $\ell$, we define, for $n\in\mathbb Z$ and $L\geq0$,
\begin{equation}\label{eq:weight}
 \phi(n)=\frac{\mathcal S_d(h(|n|-1)_+)}{2h},\quad
 \phi_L(n)=\min\{\phi(n),L\},\quad W_L=\mathrm e^{\phi_L(E_\ell)},
\end{equation}
where $x_+=\max\{x,0\}$. 
We first truncate the exponent at $L$ so that $W_L$ is bounded. Later we will consider the limit $L\to\infty$.
We write
$v_n=\one_{\{E_\ell=n\}}v$ for the component of a vector $v$ at
electric number $n$ on this link; it retains all other link degrees of
freedom.

The electric terms commute with the weight $W_L$, so we will focus on how $W_L$ affects the magnetic term $K_{\mathrm{mag}}=\sum_p[1-\cos(C\theta)_p]$.
For any real function $f:\mathbb Z\to\mathbb R$, define
\begin{equation*}
 D_f(n)=\frac{r_\ell}{2}\sum_{\sigma=\pm1}
           \bigl[\cosh(f(n+\sigma)-f(n))-1\bigr].
\end{equation*}
This nonnegative function gives an upper bound on the correction
to the magnetic quadratic form arising when we commute the weight $W=\mathrm e^{f(E_\ell)}$
past the magnetic operator. More precisely, for bounded $f$ and
$\psi\in\HH_{\mathrm{link}}$, setting $v=W\psi$ gives
\begin{equation}\label{eq:loss}
 \langle v,K_{\mathrm{mag}}v\rangle
 -\operatorname{Re}\langle W^2\psi,K_{\mathrm{mag}}\psi\rangle
 \leq\sum_n D_f(n)\|v_n\|^2.
\end{equation}
In Appendix~\ref{app:hopping}, we prove \eqref{eq:loss} first for
$\psi$ of finite electric support and then extend it by continuity
to all $\psi\in\HH_{\mathrm{link}}$.
In Appendix~\ref{app:hopping} we also show that the weight $\phi(n)$ (together with $\phi_L(n)$) in \eqref{eq:weight} satisfies
\begin{equation}\label{eq:slope-loss}
 \begin{aligned}
 |\phi(n\pm1)-\phi(n)|
 &\leq\frac12\arcosh(1+h^2n^2/(2R_d)),\\
 D_{\phi_L}(n)&\leq D_\phi(n)\leq\frac{h^2n^2}{4}.
 \end{aligned}
\end{equation}
Taking $f=\phi_L$ and $v=W_L\psi$ gives
\[
 \langle v,K_{\mathrm{mag}}v\rangle
 -\operatorname{Re}\langle W_L^2\psi,K_{\mathrm{mag}}\psi\rangle
 \leq\sum_n \frac{h^2n^2}{4}\|v_n\|^2. 
\]
In other words, the correction arising from commuting $W_L$ past $K_{\mathrm{mag}}$ is at most half of the electric energy
$h^2n^2/2$. For $\psi\in\operatorname{Dom}(K)$, combining this
estimate with Lemma~\ref{lem:local} yields
\[
\begin{aligned}
    \operatorname{Re}\langle W_L^2\psi,(K-e_0)\psi\rangle &= \operatorname{Re}\langle W_L^2\psi,K_{\mathrm{mag}}\psi\rangle + \operatorname{Re}\langle W_L^2\psi,\left(\frac{h^2}{2}\sum_{\ell'}E_{\ell'}^2\right)\psi\rangle - e_0\|v\|^2\\
    &\geq \langle v,K_{\mathrm{mag}}v\rangle + \langle v,\left(\frac{h^2}{2}\sum_{\ell'}E_{\ell'}^2\right)v\rangle - \sum_{n} \frac{h^2n^2}{4}\|v_n\|^2 -e_0\|v\|^2\\
    &=\langle v,Kv\rangle  - \sum_{n} \frac{h^2n^2}{4}\|v_n\|^2 -e_0\|v\|^2 \\
    &\geq \langle v, \left(\frac{h^2}{2}E_\ell^2 - r_\ell h\right)v \rangle- \sum_{n} \frac{h^2n^2}{4}\|v_n\|^2 \\
    &=\sum_n \left(\frac{h^2 n^2}{4}-r_\ell h\right)\|v_n\|^2.
\end{aligned}
\]
Then using $r_\ell\leq R_d$, we have
\begin{equation}\label{eq:weighted-energy}
 \operatorname{Re}\langle W_L^2\psi,(K-e_0)\psi\rangle
 \geq\sum_{n\in\mathbb Z}
       \left(\frac{h^2n^2}{4}-R_d h\right)\|(W_L\psi)_n\|^2.
\end{equation}
This inequality is the key to the electric tail bound we prove below.

\begin{proposition}[Electric tail bound]\label{prop:tail}
For every link $\ell$ and every integer $\Lambda\geq0$,
\begin{equation}\label{eq:tail}
 \left\|\one_{|E_\ell|>\Lambda}\Pi_{\EE}\right\|
 \leq q_\Lambda
 =\sqrt2\exp\left[A_d(a)
                     -\frac{\mathcal S_d(h\Lambda)}{2h}\right].
\end{equation}
The bound has no dependence on $M$, $N_p$, or the rank of $\Pi_{\EE}$.
\end{proposition}
\begin{proof}
We write $\Pi=\Pi_{\EE}$ and recall that this projection operator retains all eigenstates of
$K$ whose energy is at most $\delta=ah\EE$ above the ground energy, and filters out all other eigenstates. To obtain a bound that holds uniformly for every state in
this window, for each finite $L$ we choose a normalized
vector $\psi\in\ran\Pi$ maximizing $\|W_L\psi\|$. This is possible
because $\ran\Pi$ is finite-dimensional. With
$\lambda=\|W_L\Pi\|^2$, the maximizing vector satisfies
\[
 \Pi W_L^2\Pi\psi=\lambda\psi,\quad
 \lambda=\|W_L\psi\|^2.
\]
Since $(K-e_0)\psi\in\ran\Pi$ and $\Pi$ commutes with $K$, we obtain
\begin{equation}\label{eq:maximizer}
 \begin{aligned}
 \operatorname{Re}\langle W_L^2\psi,(K-e_0)\psi\rangle
 &=\operatorname{Re}\langle\Pi W_L^2\Pi\psi,(K-e_0)\psi\rangle\\
 &=\lambda\langle\psi,(K-e_0)\psi\rangle
 \leq\delta\lambda.
 \end{aligned}
\end{equation}
Putting $v=W_L\psi$ and using \eqref{eq:weighted-energy}, we have
\begin{equation}\label{eq:mass-energy}
 \sum_n\left(\frac{h^2n^2}{4}-R_d h-\delta\right)\|v_n\|^2\leq0.
\end{equation}
We choose $R_*=\sqrt{8(R_dh+\delta)}/h$ so that the positive
coefficient for $|n|\geq R_*$ controls the possibly negative
coefficient for $|n|<R_*$. Specifically, \eqref{eq:mass-energy} gives
\[
 0\geq -(R_dh+\delta)\sum_{|n|<R_*}\|v_n\|^2
       +(R_dh+\delta)\sum_{|n|\geq R_*}\|v_n\|^2.
\]
Since $R_dh+\delta>0$, the weighted mass outside this interval is at
most the mass inside, i.e.
\[
\sum_{|n|\geq R_*}\|v_n\|^2\leq \sum_{|n|<R_*}\|v_n\|^2.
\]
Writing $\psi_n=\one_{\{E_\ell=n\}}\psi$, we have
$v_n=\mathrm e^{\phi_L(n)}\psi_n$. To bound the weight for
$|n|<R_*$, we use
\[
\sum_{|n|<R_*}\|v_n\|^2\leq \mathrm e^{2\phi(\lceil R_*\rceil)}\sum_{|n|<R_*}\|\psi_n\|^2\leq \mathrm e^{2\phi(\lceil R_*\rceil)}.
\]
By \eqref{eq:inside-weight} in Appendix~\ref{app:mass},
\begin{equation}
    \phi(\lceil R_*\rceil)\leq \frac{2(R_d+\delta/h)}{\sqrt{R_d}}.
\end{equation}
This results in the following bound for $\|W_L\Pi\|^2=\|v\|^2\leq 2\sum_{|n|<R_*}\|v_n\|^2$:
\begin{equation*}
\|W_L\Pi\|^2\leq2\exp[\frac{4(R_d+\delta/h)}{\sqrt{R_d}}].
\end{equation*}
This operator-norm estimate bounds $\|W_L\psi\|$ for every normalized
$\psi\in\ran\Pi$, with a constant independent of $L$. We can
therefore let $L\to\infty$ by monotone convergence. Applying the
exponential Markov inequality to the spectral distribution of $E_\ell$
then gives
\[
 \|\one_{|E_\ell|>\Lambda}\psi\|^2
 \leq2\exp\left[\frac{4(R_d+\delta/h)}{\sqrt{R_d}}
                         -\frac{\mathcal S_d(h\Lambda)}h\right]
\]
for every normalized $\psi\in\ran\Pi$. 
Here we use the fact that when $|n|>\Lambda$, $2\phi(n)\geq\mathcal S_d(h\Lambda)/h$;
see
Appendix~\ref{app:markov}. Taking a square root and then the supremum
over such $\psi$ proves \eqref{eq:tail}, using additionally $\delta/h=a\EE$.
\end{proof}

\subsection{From the electric tail bound to the evolution error bound}\label{sec:dynamics}

We use the tail bound to control both the initial projection error and
the part of the exact evolution that crosses the cutoff. The latter
contribution is generated by the magnetic terms and accumulates over
the simulation time.

\begin{proof}[Proof of Theorem~\ref{thm:full}]
Write $P=P_\Lambda$, $Q=I-P$, and
$\psi(t)=\mathrm e^{-itH_a}\psi_0$. Exact evolution preserves
$\ran\Pi_{\EE}$, so Proposition~\ref{prop:tail} holds for $\psi(t)$
at every real time. Since the electric projections commute,
\begin{equation}\label{eq:union}
 Q\leq\sum_{\ell\in\mathsf L}\one_{|E_\ell|>\Lambda},\quad
 \|Q\psi(t)\|^2\leq Mq_\Lambda^2.
\end{equation}
This proves the nonzero-projection assertion when
$\sqrt Mq_\Lambda<1$.

The electric terms are diagonal and preserve the cutoff. For a
plaquette $p$, let
$\partial p=\{\ell:C_{p\ell}\neq0\}$ and define
\[
 Q_p=I-\prod_{\ell\in\partial p}\one_{|E_\ell|\leq\Lambda},
 \quad V_p=-\frac1{ah}\cos\theta_p.
\]
The operator $V_p$ acts on the four links of $\partial p$, has norm at
most $1/(ah)$, and commutes with all cutoff projections outside that
plaquette. Consequently,
\[
 PV_pQ=PV_pQ_p.
\]
Indeed, if a link outside $\partial p$ exceeds the cutoff, $V_p$
does not change its electric number and $P$ removes that component.
Thus a contribution to $PV_pQ$ must already exceed the cutoff on
one of the four links of $p$.
The four-link union bound and \eqref{eq:tail} imply
$\|Q_p\psi(t)\|\leq2q_\Lambda$. The electric and constant magnetic
terms are diagonal, so
\begin{equation*}
 \|PH_aQ\psi(t)\|
 \leq\sum_{p\in\mathsf P}\|PV_pQ_p\psi(t)\|
 \leq\frac{2N_p}{ah}q_\Lambda.
\end{equation*}

We project the exact Schr\"odinger equation onto $\ran P$ and compare
it with evolution under $H_{a,\Lambda}$. Duhamel's formula gives
\begin{equation}\label{eq:duhamel}
 P\psi(t)-\mathrm e^{-itH_{a,\Lambda}}P\psi_0
 =-i\int_0^t \mathrm e^{-i(t-s)H_{a,\Lambda}}PH_aQ\psi(s)\,ds.
\end{equation}
Unitarity bounds the norm of this difference by
$2N_p|t|q_\Lambda/(ah)$. To account for normalization of the projected
initial state, put $\beta=\|P\psi_0\|>0$. Then
\[
 \|P\psi_0-\psi_{0,\Lambda}\|
 =1-\beta\leq\sqrt{1-\beta^2}
 =\|Q\psi_0\|\leq\sqrt Mq_\Lambda.
\]
Combining this estimate with \eqref{eq:union} and
\eqref{eq:duhamel} yields
\[
 \|\psi(t)-\mathrm e^{-itH_{a,\Lambda}}\psi_{0,\Lambda}\|
 \leq\left(2\sqrt M+\frac{2N_p|t|}{ah}\right)q_\Lambda.
\]
Taking the supremum over $|t|\leq T$ proves \eqref{eq:full-error}.
\end{proof}

The tail bound remains valid throughout the exact evolution because
the Hamiltonian $H_a$ commutes with the spectral window projection $\Pi_{\mathcal{E}}$. The integral in \eqref{eq:duhamel}
therefore gives an error bound for every $T\geq0$, with the stated
linear dependence on $T$ at fixed cutoff.

\subsection{Cutoff bounds for a fixed spatial region}\label{sec:uniform}

We now derive the cutoff bounds from Theorem~\ref{thm:full}.
The spatial region, boundary conditions, and reference
scales are fixed as in Theorem~\ref{thm:general}. The link and plaquette counts
are both $\mathcal{O}(a^{-d})$. Substituting these counts and
$h=g^2$ into the definition of $L_{T,d}$ gives
\begin{equation}\label{eq:log-uniform}
 L_{T,d}\leq C_{\log}[F(a,g,T,\epsilon)+a\EE],
\end{equation}
where $C_{\log}>0$ depends only on the fixed parameters and is
independent of $g$ and $\EE$.
Appendix~\ref{app:lattice-counts} gives the counts for both boundary
conditions and an explicit choice of $C_{\log}$. This estimate separates
the dependence on $a,g,\EE,T,\epsilon$ from the constants in the cutoff bounds.

\begin{proof}[Proof of Theorem~\ref{thm:general}]
For arbitrary $h>0$, we first check that the chosen cutoff satisfies
\eqref{eq:action-target}. Put $y=\gamma_d hL_{T,d}>0$ and
$p_*=\max\{\sqrt y,y\}$. Then $\min\{p_*^2,p_*\}=y$.
By \eqref{eq:coarse-cutoff}, $h\Lambda_{\mathrm c}\geq p_*$,
so monotonicity and the lower bound \eqref{eq:global-action} give
\[
 \mathcal S_d(h\Lambda_{\mathrm c})
 \geq\mathcal S_d(p_*)\geq\frac{2y}{\gamma_d}=2hL_{T,d}.
\]
It follows that $\mathcal B_Tq_{\Lambda_{\mathrm c}}\leq\epsilon$
and $\sqrt Mq_{\Lambda_{\mathrm c}}\leq\epsilon/2<1$.
Theorem~\ref{thm:full} therefore gives a nonzero projection and
the error guarantee \eqref{eq:general-error}.

To obtain the scaling bound, we combine
\eqref{eq:coarse-cutoff} and \eqref{eq:log-uniform}:
\[
 \Lambda_{\mathrm c}
 \leq1+\frac{\sqrt{\gamma_d C_{\log}(F+a\EE)}}{g}
            +\gamma_d C_{\log}(F+a\EE)
 \leq C_d\left(\frac{\sqrt{F+a\EE}}{g}+F+a\EE\right),
\]
where $F=F(a,g,T,\epsilon)\geq1$ and one may take
$C_d=1+\sqrt{\gamma_d C_{\log}}+\gamma_d C_{\log}$.
This proves \eqref{eq:general-scaling} uniformly in $g>0$.
\end{proof}

\begin{proof}[Proof of Corollary~\ref{cor:gaussian}]
We use the quadratic lower bound \eqref{eq:quadratic-action},
proved in Appendix~\ref{app:action-bounds}. Its range
$0\leq p\leq2\sqrt{R_d}$ contains $p=h\Lambda_{G,d}$ by
the hypothesis of the corollary. Hence
\[
 \frac{\mathcal S_d(h\Lambda_{G,d})}{2h}
 \geq\frac{h\Lambda_{G,d}^2}{\gamma_d}\geq L_{T,d}.
\]
As in the preceding proof, this implies
$\mathcal B_Tq_{\Lambda_{G,d}}\leq\epsilon$ and
$\sqrt Mq_{\Lambda_{G,d}}<1$, so Theorem~\ref{thm:full} applies.
Finally, the definition of $\Lambda_{G,d}$ gives
\[
 h\Lambda_{G,d}\leq\sqrt{\gamma_d hL_{T,d}}+h.
\]
Thus \eqref{eq:general-rounding} is sufficient to ensure
$h\Lambda_{G,d}\leq2\sqrt{R_d}$.
\end{proof}

\begin{proof}[Proof of Theorem~\ref{thm:continuum}]
For $d=2$, Corollary~\ref{cor:gaussian} gives the error guarantee
and nonzero projection. To bound the cutoff, we specialize
\eqref{eq:log-uniform} to obtain $L_{T,2}\leq C_{\log}(F+a\EE)$,
where $F=F(a,g,T,\epsilon)$.
Since $\Lambda_G\geq1$, condition~\eqref{eq:gaussian-condition}
implies $g^2\leq2\sqrt2$. Together with $F+a\EE\geq1$, this gives
\[
 \Lambda_G\leq1+\frac{\sqrt{8C_{\log}(F+a\EE)}}{g}
 \leq\frac{\sqrt{2\sqrt2}+\sqrt{8C_{\log}}}{g}\sqrt{F+a\EE}.
\]
This proves \eqref{eq:main-scaling}, with the numerator as a possible
choice of $C_G$, independent of $g$ and $\EE$.
\end{proof}

\begin{proof}[Proof of Corollary~\ref{cor:polynomial}]
We write $\rho=a/a_{\refscale}$. For either boundary condition,
the lattice counts in Appendix~\ref{app:lattice-counts} give
$M=\Theta(\rho^{-2})$ and $N_p=\Theta(\rho^{-2})$.
The two terms in $\mathcal B_T$ therefore satisfy
\[
 2\sqrt M=\Theta(\rho^{-1}),\quad
 \frac{2N_pT}{ah}=\Theta(\rho^{-(4+\kappa)}).
\]
Since $\kappa\geq0$ and $T_{\refscale}>0$, the second term
dominates as $\rho\to0$. Consequently,
\[
 \log\mathcal B_T=(4+\kappa)\log(1/\rho)+\mathcal{O}(1),\quad
 \log(1/\epsilon)=\eta\log(1/\rho)+\log(1/\epsilon_{\refscale}).
\]
Since $\EE$ is fixed, $A_2(a)=2\sqrt2+\mathcal{O}(a)$. These identities give
the expansion of $L_{T,2}$ in \eqref{eq:polynomial-asymptotic}.
Because $h=e_{\mathrm{charge}}^2a_{\refscale}\rho$, we have
\[
 hL_{T,2}=\mathcal{O}(\rho\log(1/\rho))\to0,\quad
 \sqrt{8hL_{T,2}}+h\to0.
\]
Thus \eqref{eq:rounding-condition} holds for all sufficiently small
admissible $a$. Substituting the expansion of $L_{T,2}$ into
\eqref{eq:gaussian-cutoff} gives the stated cutoff asymptotic.
The quantity inside the ceiling tends to infinity, so adding less
than one does not affect this asymptotic.
\end{proof}

\section{Conclusions}

In this work, we derived improved electric-field truncation bounds for compact U(1) lattice gauge theory, with error guarantees for time evolution from states in a low-energy spectral window. The bounds improve the asymptotic dependence on the lattice spacing $a$, coupling $g$, and simulation time $T$ (exponentially in $a,T$ and polynomially in $g$), and yield substantially smaller sufficient cutoffs in the numerical comparisons in Table~\ref{tab:intro-cutoffs}. The central insight is to identify a small parameter in the Kogut--Susskind Hamiltonian and exploit tools from semiclassical analysis. Combining a local comparison of vacuum energies with Agmon's exponentially weighted energy method allows us to control electric-field tails on gauge links and translate their decay into quantitative bounds on digitization error.

Natural directions for future work include extending this approach to non-Abelian gauge theories and incorporating the resulting bounds into detailed resource estimates for quantum simulation. 
The approach developed here may also be adapted to derive error bounds
for other approximation schemes, including $q$-deformed formulations
of lattice gauge theories \cite{zache2023quantum,webbmack2026deformingtrailbaselinequantum,hayata2023q,biamonte1996SU2q}.
Another important question is whether similar improvements can be obtained for initial states with bounded local quantum numbers, as considered in Ref.~\cite{tong2022provably}. This setting differs from our assumption of a low-energy spectral window and is relevant to experimentally motivated initial states. For example, simulations of string breaking may start from a product state containing an electric flux string \cite{cao2026string}, which need not lie entirely within such a spectral window. Establishing improved truncation bounds for these initial states remains an open problem.

\section*{Acknowledgments}

This material is based upon work supported by the U.S. Department of Energy, Office of Science, Accelerated Research in Quantum Computing Centers, Quantum Utility through Advanced Computational Quantum Algorithms, Grant No. DE-SC0025572 (D.F. and Y.T.).
D.F. acknowledges the support from  National Science Foundation via the NSF CAREER award DMS-2438074.

\bibliographystyle{ieeetr}
\bibliography{ref}

\appendix

\section{Details of the Local Vacuum Comparison}\label{app:local}

We prove Lemma~\ref{lem:electric-positive} and the trial-state energy
bound \eqref{eq:trial-cost} used in Section~\ref{sec:local}. The
nonnegative Fourier coefficients of $\mu$ allow us to bound both the
derivatives of $f_s$ and the magnetic energy on the removed plaquettes.

\subsection{Nonnegative Fourier coefficients of the ground state}
\label{app:semigroup}

\begin{proof}[Proof of Lemma~\ref{lem:electric-positive}]
The operator $\overline R_\ell$ on $L^2(\TT^{M-1},d\nu_{M-1})$
has a smooth real potential and strictly positive kinetic coefficient.
For every $t>0$, $\mathrm e^{-t\overline R_\ell}$ maps each nonzero
nonnegative function to a strictly positive function. Indeed, the free
heat kernel on the connected torus is strictly positive, and the bounded
potential gives a strictly positive weight in the Feynman--Kac formula.
Together with compact resolvent, this gives a unique normalized ground
state wavefunction $\varphi_R>0$. In particular,
$\langle\varphi_R,1\rangle=\int\varphi_R\,d\nu_{M-1}>0$.

The function $1$ is constant in all $M-1$ remaining link angles and has
norm one under $d\nu_{M-1}$. Its positive overlap with $\varphi_R$
ensures that normalized imaginary-time evolution converges to
$\varphi_R$. We first show that this evolution preserves nonnegative
Fourier coefficients. For each
$p\notin S$, let $b_p\in\mathbb Z^{M-1}$ be the transpose of row $p$
of $C$ with the $\ell$th entry omitted. In the electric basis on the
remaining links,
\[
 \overline R_\ell=D_R+m_RI-A_R,\quad m_R=N_p-r_\ell,
\]
where $D_R$ represents the electric operator
$\frac{h^2}{2}\sum_{j\neq\ell}E_j^2$, and $m_RI-A_R$ represents
the magnetic operator $\sum_{p\notin S}[1-\cos\theta_p]$.
For a sequence $u=(u_n)_{n\in\mathbb Z^{M-1}}$, their actions are
\[
 (D_Ru)_n=\frac{h^2}{2}|n|^2u_n,\quad
 (A_Ru)_n=\frac12\sum_{p\notin S}(u_{n-b_p}+u_{n+b_p}).
\]
If $u_n\geq0$ for every $n$, then $(A_Ru)_n\geq0$ for every $n$.
Moreover, $\|A_R\|\leq m_R$, since each shift has norm one.
For $s\geq0$, both operators
\[
 (\mathrm e^{-sD_R}u)_n=\mathrm e^{-sh^2|n|^2/2}u_n,\quad
 \mathrm e^{-s(m_RI-A_R)}=\mathrm e^{-sm_R}\sum_{j=0}^\infty\frac{s^jA_R^j}{j!}
\]
also map nonnegative sequences to nonnegative sequences. The series
converges in operator norm because $A_R$ is bounded. The Trotter
product formula gives
\[
 \mathrm e^{-t\overline R_\ell}
 =\operatorname*{s-lim}_{J\to\infty}
    \left(\mathrm e^{-tD_R/J}\mathrm e^{-t(m_RI-A_R)/J}\right)^J
\]
in the strong operator topology. Every finite product on the right
preserves nonnegative coefficients, and convergence in $\ell^2$ implies
convergence of each coefficient. Hence $\mathrm e^{-t\overline R_\ell}$
has the same property. Since the constant function $1$ has coefficient
sequence $\one_{\{n=0\}}$, every Fourier coefficient of
$\mathrm e^{-t\overline R_\ell}1$ is nonnegative.

The simple ground eigenvalue and the positive overlap with $1$ imply,
by spectral decomposition,
\[
 \frac{\mathrm e^{-t\overline R_\ell}1}
      {\|\mathrm e^{-t\overline R_\ell}1\|}
 \longrightarrow\varphi_R\quad\text{in }L^2,
 \quad t\to\infty.
\]
Fourier coefficients are continuous linear functionals on $L^2$, so
those of $\varphi_R$ are nonnegative. We define a wavefunction on all
links by $\psi_R(\theta)=\varphi_R(\theta_{\neq\ell})$, so that it is
constant in $\theta_\ell$. Since
$R_\ell=I_\ell\otimes\overline R_\ell$, it satisfies
$R_\ell\psi_R=e_R\psi_R$. Its coefficients vanish for
$n_\ell\neq0$, and Parseval's identity gives
$\sum_n c_n^2=\|\psi_R\|^2=1$.
\end{proof}

\subsection{Estimating the energy of the trial state}\label{app:trial}

We prove the trial-energy estimate \eqref{eq:trial-cost}.
For the fixed link $\ell$, we use $S$, $C_S$, $\psi_R$, and $\mu$
from Section~\ref{sec:local}. Recall the trial factor $f_s$ and
write its square as $W_s$: for $s>0$ and $x\in\TT^{r_\ell}$,
\begin{equation*}
 W_s(x)=\sum_{z\in\mathbb Z^{r_\ell}}
                 \mathrm e^{-|x+2\pi z|^2/(2s)},\quad f_s(x)=\sqrt{W_s(x)}.
\end{equation*}
The sum and all its derivatives converge uniformly on a fundamental
cell. Thus $W_s$ is smooth and strictly positive, as is $f_s$.

To use the nonnegative Fourier coefficients of $\mu$, we first
compute those of $W_s$. Let
$Q_{\mathrm{cell}}=[-\pi,\pi)^{r_\ell}$ and use the convention
\[
 \widehat W_s(k)=\frac1{(2\pi)^{r_\ell}}
       \int_{Q_{\mathrm{cell}}}W_s(x)\mathrm e^{-ik\cdot x}\,dx,
 \quad k\in\mathbb Z^{r_\ell}.
\]
The translated cells $Q_{\mathrm{cell}}+2\pi z$, indexed by
$z\in\mathbb Z^{r_\ell}$, partition $\mathbb R^{r_\ell}$. Tonelli's
theorem and translation of each Gaussian give
\begin{align*}
 &\sum_{z\in\mathbb Z^{r_\ell}}\int_{Q_{\mathrm{cell}}}
   \left|\mathrm e^{-|x+2\pi z|^2/(2s)}\mathrm e^{-ik\cdot x}\right|dx\\
 &\quad=\sum_{z\in\mathbb Z^{r_\ell}}
          \int_{Q_{\mathrm{cell}}+2\pi z}\mathrm e^{-|y|^2/(2s)}\,dy
   =\int_{\mathbb R^{r_\ell}}\mathrm e^{-|y|^2/(2s)}\,dy
   =(2\pi s)^{r_\ell/2}<\infty.
\end{align*}
This absolute-integrability bound justifies interchanging the sum and
the Fourier integral. In its $z$th term, set $y=x+2\pi z$. The phase
is unchanged because
\[
 \mathrm e^{-ik\cdot x}=\mathrm e^{-ik\cdot y}\mathrm e^{2\pi i k\cdot z}
              =\mathrm e^{-ik\cdot y},\quad k\cdot z\in\mathbb Z.
\]
Consequently,
\begin{equation}\label{eq:wrapped-fourier}
 \begin{aligned}
 \widehat W_s(k)
 &=\frac1{(2\pi)^{r_\ell}}
   \sum_{z\in\mathbb Z^{r_\ell}}\int_{Q_{\mathrm{cell}}}
       \mathrm e^{-|x+2\pi z|^2/(2s)}\mathrm e^{-ik\cdot x}\,dx\\
 &=\frac1{(2\pi)^{r_\ell}}
   \sum_{z\in\mathbb Z^{r_\ell}}\int_{Q_{\mathrm{cell}}+2\pi z}
       \mathrm e^{-|y|^2/(2s)}\mathrm e^{-ik\cdot y}\,dy\\
 &=\frac1{(2\pi)^{r_\ell}}\int_{\mathbb R^{r_\ell}}
       \mathrm e^{-|y|^2/(2s)}\mathrm e^{-ik\cdot y}\,dy\\
 &=\left(\frac{s}{2\pi}\right)^{r_\ell/2}\mathrm e^{-s|k|^2/2}>0.
 \end{aligned}
\end{equation}
The third equality joins the translated cells. The last is the
ordinary Gaussian Fourier transform, obtained by factoring the
integral into $r_\ell$ one-dimensional Gaussian Fourier integrals.

Both the kinetic and magnetic contributions to the trial energy can
be bounded by Gaussian quadratic moments. To treat them together,
for a real symmetric positive semidefinite matrix
$A\in\mathbb R^{r_\ell\times r_\ell}$ we define
\begin{equation*}
 B_{A,s}(x)=\sum_{z\in\mathbb Z^{r_\ell}}
  (x+2\pi z)^{\mathsf T}A(x+2\pi z)
        \mathrm e^{-|x+2\pi z|^2/(2s)}.
\end{equation*}
We can again replace the integral over translated cells by an integral
over $\mathbb R^{r_\ell}$, since its absolute value is bounded by
$\|A\|\int_{\mathbb R^{r_\ell}}|y|^2\mathrm e^{-|y|^2/(2s)}\,dy$.
To compute its transform, introduce a continuous frequency
$\xi\in\mathbb R^{r_\ell}$ and set
\[
 F_s(\xi)=\frac1{(2\pi)^{r_\ell}}\int_{\mathbb R^{r_\ell}}
             \mathrm e^{-|y|^2/(2s)}\mathrm e^{-i\xi\cdot y}\,dy
 =\left(\frac{s}{2\pi}\right)^{r_\ell/2}\mathrm e^{-s|\xi|^2/2}.
\]
A Gaussian times any polynomial is integrable, so differentiating
under the integral is valid. Since
\[
 -\partial_{\xi_i}\partial_{\xi_j}F_s(\xi)
 =(s\delta_{ij}-s^2\xi_i\xi_j)F_s(\xi),
\]
where $\delta_{ij}$ is the Kronecker delta, we obtain
\begin{equation}\label{eq:moment-fourier}
 \begin{aligned}
 \widehat B_{A,s}(k)
 &=-\sum_{i,j=1}^{r_\ell}A_{ij}
     \left.\partial_{\xi_i}\partial_{\xi_j}F_s(\xi)\right|_{\xi=k}\\
 &=[s\tr A-s^2k^{\mathsf T}Ak]\widehat W_s(k)
 \leq s\tr A\,\widehat W_s(k).
 \end{aligned}
\end{equation}
Here $k^{\mathsf T}Ak\geq0$ and the strict positivity in
\eqref{eq:wrapped-fourier} justify the inequality. The coefficients
of $B_{A,s}$ themselves need not be nonnegative.

The Fourier coefficients in \eqref{eq:wrapped-fourier} and
\eqref{eq:moment-fourier} are absolutely summable. Their Fourier
series therefore converge uniformly and can be integrated term by
term against the probability measure $\mu$. Since
$\widehat\mu(-k)\geq0$ by Lemma~\ref{lem:mu-positive}, the
coefficientwise inequality \eqref{eq:moment-fourier} gives
\begin{equation}\label{eq:moment-bound}
 \begin{aligned}
 \int B_{A,s}\,d\mu
 &=\sum_{k\in\mathbb Z^{r_\ell}}
         \widehat B_{A,s}(k)\widehat\mu(-k)\\
 &\leq s\tr A\sum_{k\in\mathbb Z^{r_\ell}}
         \widehat W_s(k)\widehat\mu(-k)
 =s\tr A\int W_s\,d\mu.
 \end{aligned}
\end{equation}

We now use \eqref{eq:moment-bound} to estimate the derivatives of
$f_s$ and the magnetic energy on the plaquettes in $S$.
For fixed $x$, put $y_z=x+2\pi z$ and $w_z=\mathrm e^{-|y_z|^2/(2s)}$.
Termwise differentiation and Cauchy--Schwarz give
\begin{equation}\label{eq:gradient-bound}
 \begin{aligned}
 \nabla_x f_s(x)&=-\frac1{2s\sqrt{W_s(x)}}\sum_z y_zw_z,\\
 (\nabla_x f_s)^{\mathsf T}A\nabla_x f_s
 &=\frac{\|\sum_z w_zA^{1/2}y_z\|^2}{4s^2W_s(x)}\\
 &\leq\frac{(\sum_z w_z)(\sum_z w_z\|A^{1/2}y_z\|^2)}
                {4s^2W_s(x)}
 =\frac{B_{A,s}(x)}{4s^2}.
 \end{aligned}
\end{equation}
Both sums converge, and division is valid because $W_s(x)>0$.
Define the magnetic potential on the removed plaquettes by
$V_S(x)=\sum_{i=1}^{r_\ell}(1-\cos x_i)$. For every $z$,
periodicity and $1-\cos t\leq t^2/2$ give
$V_S(x)\leq|y_z|^2/2$. Multiplying by $w_z$ and summing yields
\begin{equation}\label{eq:cosine-bound}
 V_S(x)W_s(x)\leq\frac12 B_{I_{r_\ell},s}(x),
\end{equation}
where $I_{r_\ell}$ is the $r_\ell\times r_\ell$ identity matrix.

The trial state and its normalization are
\begin{equation*}
 \Psi_s(\theta)=Z_s^{-1/2}f_s(C_S\theta)\psi_R(\theta),\quad
 Z_s=\int_{\TT^M}W_s(C_S\theta)|\psi_R|^2\,d\nu_M
    =\int_{\TT^{r_\ell}}W_s\,d\mu>0.
\end{equation*}
To compute the electric energy, we differentiate with respect to
the $M$ link angles. Since $f_s$ is a function of the $r_\ell$
removed plaquette angles, the chain rule gives
\begin{equation}\label{eq:chain-rule}
 \nabla_\theta[f_s(C_S\theta)]
 =C_S^{\mathsf T}(\nabla_x f_s)(C_S\theta),
 \quad C_S\theta=(C\theta)_S.
\end{equation}
The gradient on the left has $M$ components; $\nabla_x$ has
$r_\ell$ components. Therefore the matrix needed in
\eqref{eq:gradient-bound} is $A=C_SC_S^{\mathsf T}$, which is
positive semidefinite and satisfies $\tr A=4r_\ell$, since every
plaquette has four distinct links with incidence coefficients
$\pm1$.

Multiplying the equation $R_\ell\psi_R=e_R\psi_R$ by
$f_s(C_S\theta)^2\psi_R$ and integrating by parts cancels the
terms containing derivatives of $\psi_R$ from the energy difference.
Appendix~\ref{app:ground-transform} gives this calculation in detail.
The resulting identity gives
\begin{equation}\label{eq:trial-energy}
 \begin{aligned}
 &\langle\Psi_s,K\Psi_s\rangle-e_R\\
 &\quad=\frac{
  \frac{h^2}{2}\int_{\TT^M}|\nabla_\theta[f_s(C_S\theta)]|^2
                           |\psi_R|^2\,d\nu_M
       +\int_{\TT^{r_\ell}}V_S(x)W_s(x)\,d\mu(x)}{Z_s}\\
 &\quad\leq\frac{h^2}{8s^2Z_s}\int B_{A,s}\,d\mu
             +\frac1{2Z_s}\int B_{I_{r_\ell},s}\,d\mu\\
 &\quad\leq\frac{h^2\tr A}{8s}+\frac{s\tr I_{r_\ell}}2
       =\frac{r_\ell h^2}{2s}+\frac{r_\ell s}{2}.
 \end{aligned}
\end{equation}
The first inequality uses \eqref{eq:chain-rule},
\eqref{eq:gradient-bound}, and \eqref{eq:cosine-bound}; the second
uses \eqref{eq:moment-bound}. This proves \eqref{eq:trial-cost}.
The choice $s=h$ gives $e_0\leq e_R+r_\ell h$ on the full link
space, whose ground energy equals the physical one as established in
Section~\ref{sec:setup}.

\subsection{Energy identity for the trial state}
\label{app:ground-transform}

We derive the cancellation used in \eqref{eq:trial-energy}.
Write
$V_R(\theta)=\sum_{p\notin S}[1-\cos(C\theta)_p]$.
The chosen ground state wavefunction is real and smooth and satisfies
$(R_\ell-e_R)\psi_R=0$ and $\partial_{\theta_\ell}\psi_R=0$.
Let $F$ be a real smooth periodic function on $\TT^M$, and write
$\partial_j=\partial_{\theta_j}$ within this calculation. The products
$F\psi_R$ and $F^2\psi_R$ are smooth and belong to the domains of
$R_\ell$ and $K$. Multiplying the ground-state equation by
$F^2\psi_R$ and integrating by parts gives
\begin{align*}
 0&=\frac{h^2}{2}\sum_{j\neq\ell}
       \int\partial_j(F^2\psi_R)\,\partial_j\psi_R\,d\nu_M
       +\int(V_R-e_R)F^2\psi_R^2\,d\nu_M\\
  &=\frac{h^2}{2}\sum_{j\neq\ell}
       \int[2F\psi_R(\partial_jF)(\partial_j\psi_R)
                  +F^2(\partial_j\psi_R)^2]\,d\nu_M
       +\int(V_R-e_R)F^2\psi_R^2\,d\nu_M.
\end{align*}
There are no boundary terms because each link angle lies on $\TT$.
On the other hand, expanding $|\partial_j(F\psi_R)|^2$ gives
\begin{align*}
 &\langle F\psi_R,(R_\ell-e_R)F\psi_R\rangle\\
 &\quad=\frac{h^2}{2}\sum_{j\neq\ell}
    \int[\psi_R^2(\partial_jF)^2
          +2F\psi_R(\partial_jF)(\partial_j\psi_R)
          +F^2(\partial_j\psi_R)^2]\,d\nu_M
      +\int(V_R-e_R)F^2\psi_R^2\,d\nu_M.
\end{align*}
Subtracting the preceding identity cancels the potential, the
cross terms, and the derivatives of the ground state wavefunction:
\begin{equation*}
 \langle F\psi_R,(R_\ell-e_R)F\psi_R\rangle
 =\frac{h^2}{2}\sum_{j\neq\ell}
     \int|\partial_jF|^2|\psi_R|^2\,d\nu_M.
\end{equation*}

We restore the removed electric and magnetic terms using
\[
 K=R_\ell+\frac{h^2}{2}E_\ell^2+V_S(C_S\theta).
\]
Because $\partial_\ell\psi_R=0$, the electric term adds
$\frac{h^2}{2}\int|\partial_\ell F|^2|\psi_R|^2\,d\nu_M$.
Adding this term to the sum over $j\neq\ell$ gives
$\frac{h^2}{2}\int|\nabla_\theta F|^2|\psi_R|^2\,d\nu_M$.
Taking $F(\theta)=f_s(C_S\theta)$ gives
\begin{align*}
 &\langle f_s(C_S\theta)\psi_R,
             (K-e_R)f_s(C_S\theta)\psi_R\rangle\\
 &\quad=\frac{h^2}{2}\int_{\TT^M}
       |\nabla_\theta[f_s(C_S\theta)]|^2|\psi_R|^2\,d\nu_M
       +\int_{\TT^M}V_S(C_S\theta)W_s(C_S\theta)
                                |\psi_R|^2\,d\nu_M.
\end{align*}
Dividing by $Z_s>0$ and applying the defining formula
\eqref{eq:mu} to the magnetic integral yields the first equality in
\eqref{eq:trial-energy}.

\section{Details of the Electric Tail Bound}\label{app:weighted}

We prove the bound \eqref{eq:loss} on the correction from commuting
the weight past $K_{\mathrm{mag}}$ and the estimates in
\eqref{eq:slope-loss}. We then justify the bound on
$\phi(\lceil R_*\rceil)$ used in Section~\ref{sec:tail} and the
application of Markov's inequality.

\subsection{Commuting the weight past the magnetic operator}
\label{app:hopping}

We express the magnetic operator as shifts in the electric basis.
For each plaquette $p$, let $\eta_p=C_{p\cdot}^{\mathsf T}\in\mathbb Z^M$.
Multiplication by $\mathrm e^{i(C\theta)_p}$ defines a unitary $T_p$
with $(T_pv)_{\mathbf m}=v_{\mathbf m-\eta_p}$ for
$\mathbf m\in\mathbb Z^M$.
Thus
\[
 K_{\mathrm{mag}}=\sum_{p\in\mathsf P}
       \left[I-\frac12(T_p+T_p^*)\right].
\]
Let $f:\mathbb Z\to\mathbb R$ be bounded, and set
$W=\mathrm e^{f(E_\ell)}$ and $v=W\psi$. We first prove
\eqref{eq:loss} when $\psi$ is a finite linear combination of
electric-basis states. Write
$v_{\mathbf m}=\langle e_{\mathbf m},v\rangle$ for the coefficient
of the configuration $\mathbf m\in\mathbb Z^M$. The diagonal terms
cancel because $\langle v,v\rangle=\langle W^2\psi,\psi\rangle$.
For each $p$, we pair the term of $T_p$ connecting $\mathbf m$ to
$\mathbf m+\eta_p$ with the reverse term of $T_p^*$. Put
\[
 d_{\mathbf m,p}=f(m_\ell+(\eta_p)_\ell)-f(m_\ell).
\]
This pair contributes
$-\operatorname{Re}(\overline{v_{\mathbf m}}v_{\mathbf m+\eta_p})$
to $\langle v,K_{\mathrm{mag}}v\rangle$. Its contribution to
$\operatorname{Re}\langle W^2\psi,K_{\mathrm{mag}}\psi\rangle$ is
\[
 \begin{aligned}
 &-\frac12\operatorname{Re}\left(
   \mathrm e^{-d_{\mathbf m,p}}
       \overline{v_{\mathbf m}}v_{\mathbf m+\eta_p}
  +\mathrm e^{d_{\mathbf m,p}}
       \overline{v_{\mathbf m+\eta_p}}v_{\mathbf m}\right)\\
 &\qquad=-\cosh(d_{\mathbf m,p})
       \operatorname{Re}(\overline{v_{\mathbf m}}v_{\mathbf m+\eta_p}).
 \end{aligned}
\]
Subtracting the second contribution from the first gives
\[
 [\cosh(d_{\mathbf m,p})-1]
       \operatorname{Re}(\overline{v_{\mathbf m}}v_{\mathbf m+\eta_p})
 \leq\frac{\cosh(d_{\mathbf m,p})-1}{2}
       (|v_{\mathbf m}|^2+|v_{\mathbf m+\eta_p}|^2).
\]
This inequality uses $\cosh t-1\geq0$ for real $t$ and
$\operatorname{Re}(\overline ab)\leq(|a|^2+|b|^2)/2$.

For $p\notin S$, $(\eta_p)_\ell=0$, so $d_{\mathbf m,p}=0$ and
the correction vanishes. For each $p\in S$, the two configurations
$\mathbf m\pm\eta_p$ have electric numbers $m_\ell\pm1$ on link
$\ell$, in one order or the other. Summing the preceding bound over
$\mathbf m$ and $p$ therefore gives the coefficient
\[
 D_f(m_\ell)=\frac{r_\ell}{2}\sum_{\sigma=\pm1}
       [\cosh(f(m_\ell+\sigma)-f(m_\ell))-1]
\]
in front of $|v_{\mathbf m}|^2$. Grouping configurations by their
electric number on link $\ell$ gives
$\|v_n\|^2=\sum_{\mathbf m:m_\ell=n}|v_{\mathbf m}|^2$, proving
\eqref{eq:loss} for $\psi$ with finite electric support.

We now extend \eqref{eq:loss} to every
$\psi\in\HH_{\mathrm{link}}$. Since $f$ is bounded,
\[
 \|W\|\leq\mathrm e^{\|f\|_\infty},\quad
 0\leq D_f(n)\leq r_\ell[\cosh(2\|f\|_\infty)-1].
\]
Also, $\|K_{\mathrm{mag}}\|\leq2N_p$. Thus both sides of
\eqref{eq:loss} are continuous functions of $\psi$ in the Hilbert-space
norm. The finite sums
\[
 \psi^{(J)}=\sum_{|\mathbf m|\leq J}
       \langle e_{\mathbf m},\psi\rangle e_{\mathbf m}
 \longrightarrow\psi
 \quad\text{in }\HH_{\mathrm{link}}
\]
therefore allow us to pass to the limit in \eqref{eq:loss}.

To include the electric energy, we take $\psi$ in the domain of $K$.
On the full link Hilbert space, boundedness of $K_{\mathrm{mag}}$ gives
\[
 \operatorname{Dom}(K)
 =\operatorname{Dom}\left(\sum_{j\in\mathsf L}E_j^2\right)
 =H^2(\TT^M).
\]
The bounded diagonal operators $W$ and $W^2$ preserve this domain.
For example, if $\psi_{\mathbf m}=\langle e_{\mathbf m},\psi\rangle$,
then
\[
 \sum_{\mathbf m}|\mathbf m|^4|(W\psi)_{\mathbf m}|^2
 \leq\mathrm e^{2\|f\|_\infty}
       \sum_{\mathbf m}|\mathbf m|^4|\psi_{\mathbf m}|^2<\infty.
\]
Because $W$ commutes with each $E_j^2$,
\[
 \operatorname{Re}\left\langle W^2\psi,
             \frac{h^2}{2}\sum_jE_j^2\psi\right\rangle
 =\left\langle v,\frac{h^2}{2}\sum_jE_j^2v\right\rangle.
\]
Rearranging \eqref{eq:loss}, adding this identity, and subtracting
$e_0\|v\|^2$ gives
\begin{equation}\label{eq:conjugated-form}
 \operatorname{Re}\langle W^2\psi,(K-e_0)\psi\rangle
 \geq\langle v,(K-e_0)v\rangle-\sum_nD_f(n)\|v_n\|^2.
\end{equation}

We next prove \eqref{eq:slope-loss} for the weights in
\eqref{eq:weight}. If $|n|\geq1$, increasing $|n|$ by one changes
the argument of $\mathcal S_d$ from $h(|n|-1)$ to $h|n|$.
Decreasing $|n|$ by one either leaves that argument at zero or gives
an interval of length $h$ closer to zero. Since $\mathcal S_d'$ is
nonnegative and increasing on $[0,\infty)$, both differences satisfy
\[
 |\phi(n\pm1)-\phi(n)|
 \leq\frac1{2h}\int_{h(|n|-1)}^{h|n|}
                \arcosh(1+s^2/(2R_d))\,ds
 \leq\frac12\arcosh(1+h^2n^2/(2R_d)).
\]
For $n=0$, both adjacent values equal zero, so the same bound holds.
For nonnegative $u,w$, we have
$|\min\{u,L\}-\min\{w,L\}|\leq|u-w|$. Hence
$|\phi_L(n\pm1)-\phi_L(n)|\leq|\phi(n\pm1)-\phi(n)|$.
Since $\cosh$ is even and increasing on $[0,\infty)$,
\begin{equation}\label{eq:D-bound}
 \begin{aligned}
 D_{\phi_L}(n)&\leq D_\phi(n)\\
 &\leq R_d\left[\cosh\left(\tfrac12\arcosh(1+h^2n^2/(2R_d))\right)-1\right]
 \leq\frac{h^2n^2}{4}.
 \end{aligned}
\end{equation}
The last step uses
$\cosh(x/2)-1\leq(\cosh x-1)/2$, which follows from convexity,
with $x=\arcosh(1+h^2n^2/(2R_d))$. This proves that the correction
is at most half of the electric energy $h^2n^2/2$ on link $\ell$.

For finite $L$, we can apply \eqref{eq:conjugated-form} with
$f=\phi_L$ to every $\psi\in\operatorname{Dom}(K)$. Combining it
with \eqref{eq:local} and \eqref{eq:D-bound}, and writing
$v=W_L\psi$, gives
\[
 \begin{aligned}
 \operatorname{Re}\langle W_L^2\psi,(K-e_0)\psi\rangle
 &\geq\sum_n\left(\frac{h^2n^2}{2}-r_\ell h
                         -D_{\phi_L}(n)\right)\|v_n\|^2\\
 &\geq\sum_n\left(\frac{h^2n^2}{4}-r_\ell h\right)\|v_n\|^2\\
 &\geq\sum_n\left(\frac{h^2n^2}{4}-R_dh\right)\|v_n\|^2.
 \end{aligned}
\]
This proves \eqref{eq:weighted-energy} on the domain of $K$.
In particular, it applies to every $\psi\in\ran\Pi_{\EE}$, since
$\Pi_{\EE}$ projects onto a bounded interval of eigenvalues of $K$.

\subsection{Bounding the weighted norm}
\label{app:mass}

We give the coefficient bounds and the estimate of
$\phi(\lceil R_*\rceil)$ used in Section~\ref{sec:tail}. Let
$\psi\in\ran\Pi_{\EE}$ be the normalized vector maximizing
$\|W_L\psi\|$, as chosen before \eqref{eq:maximizer}, and put
$v=W_L\psi$. Recall that $\delta=ah\EE\geq0$ and
\begin{equation*}
 R_*^2=\frac{8(R_d h+\delta)}{h^2}>0.
\end{equation*}
The coefficient in \eqref{eq:mass-energy} satisfies
\begin{equation*}
 \frac{h^2n^2}{4}-R_d h-\delta
 \geq\begin{cases}
  -(R_d h+\delta),&|n|<R_*,\\
  R_d h+\delta,&|n|\geq R_*.
 \end{cases}
\end{equation*}
The first bound uses $n^2\geq0$, and the second uses
$n^2\geq R_*^2$. Splitting the sum into these two sets gives
\begin{equation*}
 \begin{aligned}
 0&\geq\sum_{n\in\mathbb Z}
       \left(\frac{h^2n^2}{4}-R_d h-\delta\right)\|v_n\|^2\\
 &=\sum_{|n|<R_*}\left(\frac{h^2n^2}{4}-R_d h-\delta\right)\|v_n\|^2
   +\sum_{|n|\geq R_*}\left(\frac{h^2n^2}{4}-R_d h-\delta\right)\|v_n\|^2\\
 &\geq-(R_d h+\delta)\sum_{|n|<R_*}\|v_n\|^2
       +(R_d h+\delta)\sum_{|n|\geq R_*}\|v_n\|^2.
 \end{aligned}
\end{equation*}
Since $R_d h+\delta>0$, rearranging and dividing gives
\begin{equation}\label{eq:mass-comparison}
 \sum_{|n|\geq R_*}\|(W_L\psi)_n\|^2
 \leq\sum_{|n|<R_*}\|(W_L\psi)_n\|^2.
\end{equation}

To bound the weight for $|n|<R_*$, we use
$\arsinh u=\int_0^u(1+t^2)^{-1/2}\,dt\leq u$ for $u\geq0$.
For every $p\geq0$, this gives
\[
 \mathcal S_d(p)
 =2\int_0^p\arsinh\left(\frac{s}{2\sqrt{R_d}}\right)\,ds
 \leq\int_0^p\frac{s}{\sqrt{R_d}}\,ds
 =\frac{p^2}{2\sqrt{R_d}}.
\]
Since $R_*>0$, we have $(\lceil R_*\rceil-1)_+\leq R_*$. Thus
\begin{equation}\label{eq:inside-weight}
 \begin{aligned}
 \phi(\lceil R_*\rceil)
 &=\frac{\mathcal S_d(h(\lceil R_*\rceil-1)_+)}{2h}\\
 &\leq\frac{h(\lceil R_*\rceil-1)_+^2}{4\sqrt{R_d}}
 \leq\frac{hR_*^2}{4\sqrt{R_d}}
 =\frac{2(R_d+\delta/h)}{\sqrt{R_d}}.
 \end{aligned}
\end{equation}
Write $\psi_n=\one_{\{E_\ell=n\}}\psi$, so that
$v_n=\mathrm e^{\phi_L(n)}\psi_n$. For $|n|<R_*$,
monotonicity in $|n|$ gives
$\phi_L(n)\leq\phi(n)\leq\phi(\lceil R_*\rceil)$.
Combining this bound with \eqref{eq:mass-comparison}, we obtain
\begin{equation}\label{eq:mass-to-norm}
 \begin{aligned}
 \|W_L\Pi_{\EE}\|^2
 &=\|W_L\psi\|^2\\
 &=\sum_{|n|<R_*}\|v_n\|^2+\sum_{|n|\geq R_*}\|v_n\|^2\\
 &\leq2\sum_{|n|<R_*}\|v_n\|^2\\
 &\leq2\mathrm e^{2\phi(\lceil R_*\rceil)}
                       \sum_{|n|<R_*}\|\psi_n\|^2\\
 &\leq2\mathrm e^{2\phi(\lceil R_*\rceil)}
 \leq2\exp[\frac{4(R_d+\delta/h)}{\sqrt{R_d}}].
 \end{aligned}
\end{equation}
The first equality uses the maximizing property of $\psi$. The last
line uses $\sum_n\|\psi_n\|^2=\|\psi\|^2=1$ and
\eqref{eq:inside-weight}.

\subsection{From the weighted norm to the electric tail bound}
\label{app:markov}

We now take any normalized $\psi\in\ran\Pi_{\EE}$ and write
$\psi_n=\one_{\{E_\ell=n\}}\psi$. The operator-norm bound in
\eqref{eq:mass-to-norm} gives, for every finite $L\geq0$,
\[
 \|W_L\psi\|^2\leq\|W_L\Pi_{\EE}\|^2
 \leq2\exp\left[\frac{4(R_d+\delta/h)}{\sqrt{R_d}}\right].
\]
The bound is independent of $L$. Since
$\phi_L(n)=\min\{\phi(n),L\}$ increases to $\phi(n)$ for each $n$,
monotone convergence gives
\begin{equation*}
 \begin{aligned}
 \sum_{n\in\mathbb Z}\mathrm e^{2\phi(n)}\|\psi_n\|^2
 &=\lim_{L\to\infty}\sum_{n\in\mathbb Z}
           \mathrm e^{2\phi_L(n)}\|\psi_n\|^2\\
 &=\lim_{L\to\infty}\|W_L\psi\|^2
 \leq2\exp\left[\frac{4(R_d+\delta/h)}{\sqrt{R_d}}\right].
 \end{aligned}
\end{equation*}
The nonnegative numbers $\|\psi_n\|^2$ sum to one and are the
spectral probability distribution of $E_\ell$ in the state $\psi$.
For an integer $\Lambda\geq0$, $|n|>\Lambda$ implies
$|n|\geq\Lambda+1$, so
\[
 2\phi(n)\geq2\phi(\Lambda+1)=\frac{\mathcal S_d(h\Lambda)}h.
\]
Applying Markov's inequality to $\mathrm e^{2\phi(n)}$ under this
distribution gives
\begin{equation*}
 \begin{aligned}
 \|\one_{|E_\ell|>\Lambda}\psi\|^2
 &=\sum_{|n|>\Lambda}\|\psi_n\|^2\\
 &\leq \mathrm e^{-\mathcal S_d(h\Lambda)/h}
            \sum_{|n|>\Lambda}\mathrm e^{2\phi(n)}\|\psi_n\|^2\\
 &\leq2\exp\left[\frac{4(R_d+\delta/h)}{\sqrt{R_d}}
                       -\frac{\mathcal S_d(h\Lambda)}h\right].
 \end{aligned}
\end{equation*}
Taking a square root and then the supremum over normalized
$\psi\in\ran\Pi_{\EE}$, and using $\delta/h=a\EE$, gives
\[
 \|\one_{|E_\ell|>\Lambda}\Pi_{\EE}\|
 \leq\sqrt2\exp\left[A_d(a)
                    -\frac{\mathcal S_d(h\Lambda)}{2h}\right].
\]
This is the bound in Proposition~\ref{prop:tail}.

\section{Lattice Counts and Cutoff Estimates}
\label{app:action}

We give the lattice counts and bounds on $\mathcal S_d$ used in
Section~\ref{sec:uniform}, then examine how the cutoff depends on
simulation time and lattice spacing. Throughout this appendix, the
spatial dimension $d$ is fixed.

\subsection{Lattice counts and the bound on \texorpdfstring{$L_{T,d}$}{L(T,d)}}
\label{app:lattice-counts}

We prove \eqref{eq:log-uniform} by counting the links and plaquettes
in the fixed spatial region. With periodic boundaries, a link is
specified by its starting vertex and direction, and a plaquette by
its base vertex and a pair of directions. Hence
\begin{equation}\label{eq:counts-periodic}
 M=\frac{d\mathcal V}{a^d},\quad
 N_p=\binom d2\frac{\mathcal V}{a^d}.
\end{equation}
With open boundaries, a link has $N_i$ possible positions along its
direction $i$ and $N_k+1$ vertex positions in each transverse direction
$k$. A plaquette has $N_iN_j$ positions in its plane and $N_k+1$
positions in each remaining direction. Thus
\begin{equation}\label{eq:counts}
 M=\sum_{i=1}^d N_i\prod_{k\neq i}(N_k+1),\quad
 N_p=\sum_{1\leq i<j\leq d}N_iN_j\prod_{k\neq i,j}(N_k+1).
\end{equation}
Products over an empty set equal one. Substituting $N_i=L_i/a$
and using $a\leq a_{\refscale}$, we define
\[
 C_M=\sum_{i=1}^d L_i\prod_{k\neq i}(L_k+a_{\refscale}),\quad
 C_P=\sum_{1\leq i<j\leq d}L_iL_j
                          \prod_{k\neq i,j}(L_k+a_{\refscale}).
\]
These constants depend only on the fixed spatial region, dimension,
and reference spacing. For either boundary condition,
\[
 M\leq C_Ma^{-d},\quad N_p\leq C_Pa^{-d}.
\]

Let $\rho=a/a_{\refscale}\in(0,1]$ and
$\tau=T/T_{\refscale}\geq0$. With
$h=g^2$,
\begin{equation}\label{eq:prefactor-uniform}
 \begin{aligned}
 \mathcal B_T
 &\leq
 \frac{2\sqrt{C_M}}{a_{\refscale}^{d/2}}\rho^{-d/2}
 +\frac{2C_PT_{\refscale}}{a_{\refscale}^{d+1}}
                 g^{-2}\tau\rho^{-(d+1)}\\
 &\leq C_B\rho^{-b_d}(1+\tau)(1+g^{-2}),
 \end{aligned}
\end{equation}
where the fixed dimensionless constants are
\[
 b_d=d+1,\quad
 C_B=\frac{2\sqrt{C_M}}{a_{\refscale}^{d/2}}
       +\frac{2C_PT_{\refscale}}
                   {a_{\refscale}^{d+1}}.
\]
The power $d+1$ follows from the plaquette count
$N_p=\mathcal{O}(a^{-d})$ and the factor $1/a$ in $\mathcal B_T$.
We retain the term $2a\EE/\sqrt{R_d}$ in $A_d(a)$ explicitly and set
\[
 C_0=2\sqrt{R_d}+|\log(\sqrt2C_B)|,\quad
 C_{\log}=\max\{C_0,b_d,1,2/\sqrt{R_d}\}.
\]
Taking logarithms of \eqref{eq:prefactor-uniform} and adding $A_d(a)$ yields
\begin{equation*}
 \begin{aligned}
 L_{T,d}
 &\leq C_0+b_d\log(1/\rho)+\log(1+\tau)+\log(1/\epsilon)
                         +\log(1+g^{-2})+\frac{2a\EE}{\sqrt{R_d}}\\
 &\leq C_{\log}[F(a,g,T,\epsilon)+a\EE].
 \end{aligned}
\end{equation*}

This proves \eqref{eq:log-uniform}, with constants independent of
$a,g,\EE,T,\epsilon$.
For the fixed-bare-charge family, let
$c=(e_{\mathrm{charge}}^2a_{\refscale}^{3-d})^{-1}$. Then
$g^{-2}=c\rho^{d-3}$ and
\[
 \log(1+g^{-2})\leq\log(1+c)+\max\{3-d,0\}\log(1/\rho).
\]

\subsection{Bounds for \texorpdfstring{$\mathcal{S}_d$}{S(d)}}\label{app:action-bounds}

For $p\geq0$, differentiation of \eqref{eq:action} gives
\[
 \mathcal S_d'(p)
 =\arcosh\left(1+\frac{p^2}{2R_d}\right)
 =2\arsinh\frac{p}{2\sqrt{R_d}},\quad
 \mathcal S_d''(p)=\frac{2}{\sqrt{p^2+4R_d}}>0.
\]
Thus $\mathcal S_d$ is increasing and convex on $[0,\infty)$.
To obtain the quadratic bound used in Corollary~\ref{cor:gaussian},
we use, for $0\leq u\leq1$,
\[
 \frac{u}{\sqrt2}\leq
 \arsinh u=\int_0^u\frac{dt}{\sqrt{1+t^2}}\leq u.
\]
Substituting $u=s/(2\sqrt{R_d})$ and integrating over $0\leq s\leq p$
shows that
\begin{equation}\label{eq:quadratic-action}
 \frac{p^2}{2\sqrt{2R_d}}
   =\frac{2p^2}{\gamma_d}
 \leq\mathcal S_d(p)\leq\frac{p^2}{2\sqrt{R_d}},
 \quad 0\leq p\leq2\sqrt{R_d}.
\end{equation}
The upper bound holds for every $p\geq0$, since $\arsinh u\leq u$
does not require $u\leq1$.

To obtain a lower bound at all $p$, use $R_d\geq2$, so that
\eqref{eq:quadratic-action} applies on $[0,1]$. For $p\geq1$,
convexity and $\mathcal S_d(0)=0$ imply
$\mathcal S_d(p)/p\geq\mathcal S_d(1)\geq2/\gamma_d$.
Combining the quadratic bound for $0\leq p\leq1$ with this linear
bound for $p\geq1$ proves \eqref{eq:global-action}, which is used
for the cutoff in Theorem~\ref{thm:general}.

\subsection{Inverting \texorpdfstring{$\mathcal S_d$}{S(d)}}\label{app:inverse}

The cutoff in \eqref{eq:full-cutoff} uses $\mathcal S_d$ directly
and can improve on cutoffs obtained from its lower bounds. We examine
its dependence on $hL_{T,d}$. For this subsection, write $L=L_{T,d}$
and define the real-valued cutoff before taking the ceiling:
\[
 \lambda_*=\frac{\mathcal S_d^{-1}(2hL)}h.
\]
By \eqref{eq:full-cutoff},
$\Lambda_{\mathrm{full}}=\lceil\lambda_*\rceil$.
We retain the distinction between $\lambda_*$ and its ceiling when
the limiting value is finite.

As $p\to0$, the Taylor expansion of $\arsinh$ in
\eqref{eq:action} gives
\[
 \mathcal S_d(p)
 =\frac{p^2}{2\sqrt{R_d}}
   -\frac{p^4}{96R_d^{3/2}}+\mathcal{O}_d(p^6).
\]
Consequently, when $hL\to0$,
\begin{equation*}
 \lambda_*=
 \sqrt{\frac{4\sqrt{R_d}\,L}{h}}\,[1+\mathcal{O}_d(hL)].
\end{equation*}
Compared with $\sqrt{\gamma_dL/h}$ in
\eqref{eq:general-gaussian}, this has the same dependence on $L/h$
and a smaller leading coefficient.
When $hL$ is bounded above and below by positive constants,
$\mathcal S_d^{-1}(2hL)$ is also bounded above and below by positive
constants. Hence $\lambda_*=\Theta_d(1/h)$ and the integer cutoff
is $\Theta_d(1+1/h)$, with constants also depending on those bounds.

As $p\to\infty$, the formula in \eqref{eq:action} gives
\begin{equation}\label{eq:large-action}
 \mathcal S_d(p)
 =2p\log\frac{p}{\mathrm e\sqrt{R_d}}
       +4\sqrt{R_d}+\mathcal{O}_d(p^{-1}).
\end{equation}
Here $\mathrm e$ is Euler's number. To obtain an explicit sufficient
cutoff without evaluating $\mathcal S_d^{-1}$, we also use a lower
bound valid for every $p>0$. Since
$\arsinh u=\log(u+\sqrt{1+u^2})\geq\log(2u)$ for $u>0$
and the logarithm is integrable at zero, integration gives
\begin{equation}\label{eq:large-lower}
 \mathcal S_d(p)
 \geq2p\log\frac{p}{\mathrm e\sqrt{R_d}},
 \quad p>0.
\end{equation}

For $z>0$, let $W_0(z)$ be the unique positive solution of
$W_0(z)\exp(W_0(z))=z$, the principal real branch of Lambert's
function. With
$w=W_0(hL/(\mathrm e\sqrt{R_d}))$ and $p=hL/w$,
we have $p/(\mathrm e\sqrt{R_d})=\exp(w)$.
Thus the right-hand side of \eqref{eq:large-lower} equals $2hL$,
which ensures \eqref{eq:action-target}. An explicit sufficient
integer cutoff is therefore
\begin{equation}\label{eq:lambert-cutoff}
 \Lambda_W=
 \left\lceil\frac{L_{T,d}}
 {W_0(hL_{T,d}/(\mathrm e\sqrt{R_d}))}\right\rceil.
\end{equation}
The lower bound \eqref{eq:large-lower} is loose at small $p$, so
\eqref{eq:lambert-cutoff} is mainly useful when $hL_{T,d}$ is large.

For the asymptotic behavior of the inverse, \eqref{eq:large-action}
implies
$\mathcal S_d^{-1}(x)\sim x/(2\log x)$ as $x\to\infty$.
Indeed, for $p=x/(2\log x)$ and any fixed $0<\zeta<1$,
\eqref{eq:large-action} gives
$\mathcal S_d((1\pm\zeta)p)/x\to1\pm\zeta$. Thus, for sufficiently
large $x$,
\[
 \mathcal S_d((1-\zeta)p)<x<\mathcal S_d((1+\zeta)p).
\]
Since $\mathcal S_d$ is strictly increasing on $[0,\infty)$,
\[
 (1-\zeta)p<\mathcal S_d^{-1}(x)<(1+\zeta)p.
\]
Dividing by $p$ and then letting $\zeta\to0$ proves the inverse
asymptotic. Substituting $x=2hL_{T,d}$ gives
\begin{equation}\label{eq:large-inverse}
 \lambda_*\sim\frac{L_{T,d}}{\log(hL_{T,d})},
 \quad hL_{T,d}\to\infty.
\end{equation}
This asymptotic is for $\lambda_*$. It also holds for
$\Lambda_{\mathrm{full}}$ if the right-hand side diverges.
In all cases $\lambda_*\leq\Lambda_{\mathrm{full}}<\lambda_*+1$.

\subsection{Exact dependence on simulation time}\label{app:time}

To make the dependence on simulation time explicit, we define
\begin{equation*}
 T_c=\frac{ah\sqrt M}{N_p}.
\end{equation*}
The quantity $T_c$ is positive because the lattice contains
plaquettes. At $T=T_c$, the two terms in $\mathcal B_T$ are equal.
Factoring out $2\sqrt M$ gives
\begin{equation}\label{eq:time-target}
 \mathcal B_T=2\sqrt M(1+T/T_c),\quad
 L_{T,d}=A_d(a)+\log\frac{2\sqrt{2M}}{\epsilon}
                         +\log(1+T/T_c).
\end{equation}
For fixed lattice, coupling, $\EE$, and $\epsilon$, the dependence
of $L_{T,d}$ on time is entirely through $\log(1+T/T_c)$. If
$\lambda_{G,d}(T)=\sqrt{\gamma_dL_{T,d}/h}$ denotes the
value in \eqref{eq:general-gaussian} before taking the ceiling, then
\[
 \lambda_{G,d}(T_2)^2-\lambda_{G,d}(T_1)^2
 =\frac{\gamma_d}{h}
       \log\frac{1+T_2/T_c}{1+T_1/T_c}.
\]
The error guarantee from Corollary~\ref{cor:gaussian} requires
$h\lceil\lambda_{G,d}(T)\rceil\leq2\sqrt{R_d}$.
For arbitrarily large $T$ at fixed lattice, coupling, $\EE$, and
$\epsilon$, we use \eqref{eq:large-inverse} instead. It gives
\[
 \Lambda_{\mathrm{full}}
 =\mathcal{O}\left(\frac{\log(1+T/T_c)}
                    {\log\log(1+T/T_c)}\right),
 \quad T/T_c\to\infty.
\]

If we keep the integer cutoff fixed and solve
$\mathcal B_Tq_\Lambda\leq\epsilon$ for $T$, we obtain
\[
 0\leq T\leq\frac{ah}{2N_p}
 \left[
   \frac{\epsilon}{\sqrt2}
       \exp\left(\frac{\mathcal S_d(h\Lambda)}{2h}-A_d(a)\right)
       -2\sqrt M
 \right],
\]
provided the bracket is nonnegative.
Throughout this interval, Theorem~\ref{thm:full} guarantees error at
most $\epsilon$ for every normalized initial state in
$\ran\Pi_{\EE}$.

For periodic boundaries at fixed bare charge,
\eqref{eq:counts-periodic} gives the exact formulas
\begin{equation*}
 \mathcal B_T=
 \frac{2\sqrt{d\mathcal V}}{a^{d/2}}
 +\frac{2\binom d2\mathcal V T}{e_{\mathrm{charge}}^2a^4},
 \quad
 T_c=\frac{e_{\mathrm{charge}}^2\sqrt d}
              {\binom d2\sqrt{\mathcal V}}\,a^{4-d/2}.
\end{equation*}
For $d=2$, writing the area as $\mathcal A=L_1L_2$,
\eqref{eq:time-target} becomes
\[
 L_{T,2}
 =\sqrt2(2+a\EE)
  +\log\frac{4\sqrt{\mathcal A}}{a\epsilon}
  +\log\left(1+
       \frac{\sqrt{\mathcal A}\,T}
            {\sqrt2\,e_{\mathrm{charge}}^2a^3}\right).
\]
For fixed $\EE$ and $\epsilon$ as $a\to0$, this formula gives
$L_{0,2}=\log(a_{\refscale}/a)+\mathcal{O}(1)$ at zero time, whereas at any
fixed positive time $L_{T,2}=4\log(a_{\refscale}/a)+\mathcal{O}(1)$.

\subsection{Decreasing the lattice spacing at fixed bare charge}
\label{app:refinement}

We consider the fixed-bare-charge case of Theorem~\ref{thm:general}, with
the spatial region, boundary conditions, and reference scales fixed. We also fix
$\EE\geq0$ independently of $a$. We consider an admissible
sequence $a\to0$ and write $\rho=a/a_{\refscale}$.
For either boundary condition, the exact counts
\eqref{eq:counts-periodic}--\eqref{eq:counts} imply
$M=\Theta(a^{-d})$ and $N_p=\Theta(a^{-d})$.
If
\[
 T=T_{\refscale}\rho^{-\kappa},\quad
 \epsilon=\epsilon_{\refscale}\rho^\eta,\quad
 \kappa,\eta\geq0,\quad 0<\epsilon_{\refscale}<1,
\]
then the two terms in $\mathcal B_T$ satisfy
\[
 2\sqrt M=\Theta(\rho^{-d/2}),\quad
 \frac{2N_pT}{ah}=\Theta(\rho^{-(4+\kappa)}).
\]
Consequently,
\begin{equation}\label{eq:dimension-target}
 L_{T,d}=\alpha_d\log(1/\rho)+\mathcal{O}(1),\quad
 \alpha_d=\max\{d/2,4+\kappa\}+\eta.
\end{equation}
The maximum in $\alpha_d$ accounts for which of the two terms
in $\mathcal B_T$ dominates.

For $d=2$, $h=e_{\mathrm{charge}}^2a$ and $hL_{T,2}\to0$.
This is the setting of Corollary~\ref{cor:polynomial}.
Thus $h\Lambda_G\leq2\sqrt2$ holds for all sufficiently small
admissible $a$, and
\[
 \Lambda_G
 =\mathcal{O}\left(a^{-1/2}\sqrt{\log(1/\rho)}\right).
\]
At $T=0$ with the same error scaling, $\mathcal B_0=2\sqrt M$
and $L_{0,2}=(1+\eta)\log(1/\rho)+\mathcal{O}(1)$.
For an example with faster growth of time, take
$T=T_{\refscale}\exp(c/\rho)$ with $c>0$ and fixed $\epsilon$.
Then $L_{T,2}\sim c/\rho$ and
$hL_{T,2}\to c e_{\mathrm{charge}}^2a_{\refscale}$.
The cutoff $\Lambda_{\mathrm{full}}$ is $\Theta(a^{-1})$ in this
case. Whether $h\Lambda_G\leq2\sqrt2$ holds depends on the
limiting constant, whereas Theorem~\ref{thm:full} applies for every
$c>0$.

For $d=3$, $h=e_{\mathrm{charge}}^2$ stays fixed and
$hL_{T,3}\to\infty$. Consequently, $h\Lambda_{G,3}$ diverges,
so the condition in Corollary~\ref{cor:gaussian} eventually fails
along the polynomial family above.
Since $\alpha_3=4+\kappa+\eta$, \eqref{eq:large-inverse} gives
\[
 \Lambda_{\mathrm{full}}
 \sim\frac{(4+\kappa+\eta)\log(1/\rho)}
                  {\log\log(1/\rho)}.
\]
The right-hand side diverges, so rounding does not change this
asymptotic.

For $d>3$, $h=e_{\mathrm{charge}}^2a^{3-d}$ grows and
\[
 \log(hL_{T,d})=(d-3)\log(1/\rho)
                         +\mathcal{O}(\log\log(1/\rho)).
\]
Combining this identity with \eqref{eq:dimension-target}
and \eqref{eq:large-inverse} yields
\[
 \lambda_*\longrightarrow\frac{\alpha_d}{d-3},\quad
 \Lambda_{\mathrm{full}}=\mathcal{O}(1).
\]
Here the ceiling cannot be omitted because $\lambda_*$ has a finite
limit. Along this family with fixed bare charge, the cutoff obtained by
inverting $\mathcal S_d$ is bounded, while Theorem~\ref{thm:general}
gives the upper bound $\Lambda_{\mathrm c}=\mathcal{O}(\log(1/\rho))$.

\section{Truncation bound computation for \texorpdfstring{Ref.~\cite{tong2022provably}}{Ref 10}}

\label{sec:truncation_bound_tong_etal}

To evaluate the dynamical estimate of Ref.~\cite{tong2022provably} on the same initial
states as other methods in Table~\ref{tab:intro-cutoffs}, we first choose an initial cutoff $\Lambda_0$ using
\eqref{eq:full-cutoff} at time zero and accuracy $\epsilon/4$.
Thus this estimate for Ref.~\cite{tong2022provably} uses the static estimate for initialization in this work.
Set $x=\chi T$, with $\chi=R_d/(ag^2)$ as defined in Ref.~\cite{tong2022provably}. For an integer $k\geq1$
with $k+1>x$, the first inequality in
Ref.~\cite[Eq.~(44)]{tong2022provably} bounds the norm of the tail on each
link beyond $\Lambda_0+k-1$ by
\begin{equation}\label{eq:intro-tong-tail}
 \sum_{j=k}^{\infty}\frac{x^j}{j!}
 \leq b_k(x):=\frac{x^k}{k!\,[1-x/(k+1)]}.
\end{equation}
Here the Dyson-series estimate is valid for arbitrary $T$ because the
hopping operator is bounded. We choose the smallest such $k$ satisfying
\begin{equation}\label{eq:intro-tong-cutoff}
 \mathcal B_T b_k(x)\leq\frac\epsilon2.
\end{equation}
The truncation is therefore
\[
\Lambda_{(2)}=\Lambda_0+k-1.
\]
The union bound and Duhamel argument in Section~\ref{sec:dynamics}
then bound the dynamical error from the normalized $\Lambda_0$ projection
by $\mathcal B_Tb_k(x)$. Nesting of the cutoff spaces bounds both its
distance to the original state and its distance to the normalized
projection by $\epsilon/4$.
The total error is therefore at most $\epsilon$.

\end{document}